\documentclass[aps,prc,twocolumn,floatfix,showpacs,preprintnumbers,amsmath,amssymb,nofootinbib,groupedaddress]{revtex4-2}
\usepackage{color}
\usepackage{graphicx}
\graphicspath{{figures/}} 
\usepackage{dcolumn}    
\usepackage{multirow}
\usepackage{bm}         
\usepackage{sidecap}
\usepackage{hyperref}   
\usepackage{amsmath}
\usepackage{graphicx}
\usepackage{float}
\usepackage{adjustbox}
\usepackage{caption}
\usepackage{subcaption}

\usepackage{mathtools} 
\usepackage[version=3]{mhchem} 

\usepackage[dvipsnames,usenames]{xcolor}
\usepackage{mathrsfs,natbib}
\usepackage{epsf,amssymb,amsbsy,amsfonts,amssymb,amsmath}
\usepackage{slashed}
\usepackage{comment}

\usepackage{hyperref}
\usepackage[justification=raggedright]{caption}
\definecolor{dark-red}{rgb}{0.,0.,0}
\definecolor{dark-blue}{rgb}{0.,0.,1}
\definecolor{medium-blue}{rgb}{0,0,1}
\hypersetup{
    colorlinks, linkcolor={dark-red},
    citecolor={dark-blue}, urlcolor={medium-blue}
}

\newcommand{\sat}{\mathrm{sat}}
\newcommand{\sym}{\mathrm{sym}}

\newcommand{\NM}{\mathrm{NM}}

\newcommand{\fopt}{\mathrm{PT}}

\begin{document}
%
\title{On the nature of compact stars determined by gravitational waves, radio-astronomy, x-ray emission and nuclear physics}

\author{H. G\"{u}ven}
\affiliation{Universit\'e Paris-Saclay, CNRS/IN2P3, IJCLab, 91405 Orsay, France}
\affiliation{Physics Department, Yildiz Technical University, 34220 Esenler, Istanbul, Turkey}

\author{J. Margueron}
\affiliation{Univ Lyon, Univ Claude Bernard Lyon 1, CNRS/IN2P3, IP2I Lyon, UMR 5822, F-69622, Villeurbanne, France}
\date{\today}

\author{K. Bozkurt}
\affiliation{Physics Department, Yildiz Technical University, 34220 Esenler, Istanbul, Turkey}
\affiliation{Universit\'e Paris-Saclay, CNRS/IN2P3, IJCLab, 91405 Orsay, France}

\author{E. Khan}
\affiliation{Universit\'e Paris-Saclay, CNRS/IN2P3, IJCLab, 91405 Orsay, France}
\affiliation{Institut Universitaire de France (IUF)}

%
\begin{abstract}
We investigate the question of the nature of compact stars, considering they may be neutron stars or hybrid stars containing a quark core, within the present constraints given by gravitational waves, radio-astronomy, X-ray emissions from millisecond pulsars and nuclear physics. A Bayesian framework is used to combine together all these constraints and to predict tidal deformabilities and radii for a 1.4~M$_\odot$ compact star.
We find that present gravitation wave and radio-astronomy data favors stiff nucleonic EoS compatible with nuclear physics and that GW170817 
waveform is best described for binary hybrid stars. Binary neutron stars with soft EoS could however not be totally excluded.
In all cases, these 
data favor stiff quark matter, independently of the nuclear EoS, with a low value for the transition density ($n_\mathrm{tr}\in[0.18,0.35]~\mathrm{fm}^{-3}$).
Combining these results with constraints from X-ray observation supports the existence $1.4$~M$_\odot$ mass hybrid star, with a radius predicted to be about $R_{1.4}=12.22(45)$~km. 
\end{abstract}

\maketitle

Neutron Stars (NSs), and most generally compact stars (CSs), are forefront laboratories to explore the properties of extreme matter and consequently the strong interaction. The guidance of experimental and observational data are of prime importance, especially because the theory of the strong interaction, the quantum chromo-dynamics (QCD), could not be simply applied in the regime explored by CSs. One of the most important question in this field is to understand how the strong interaction evolves as function of the density and how quark matter emerges from hadrons~\cite{Alford:2008,Anglani:2014}, if it ever does for densities relevant for stable CSs. Therefore, one should consider two types of CS: NS, with no phase transition in the core, and hybrid stars (HS), with phase transition, towards quark matter, in the core. Note that the existence of strange quark stars is not considered in our analysis since we assume that matter is ruled by a single equation of state. In the present paper, we investigate the onset of a first order phase transition (FOPT) producing the largest correction to the global properties of CSs. It should be noted that CSs observations are also complementary to Earth experiments, such as for instance heavy ion collision~\cite{HADES:2019}, since they explore very isospin asymmetric and dense matter.

\begin{figure}[t]
\centering
\includegraphics[width=0.5\textwidth]{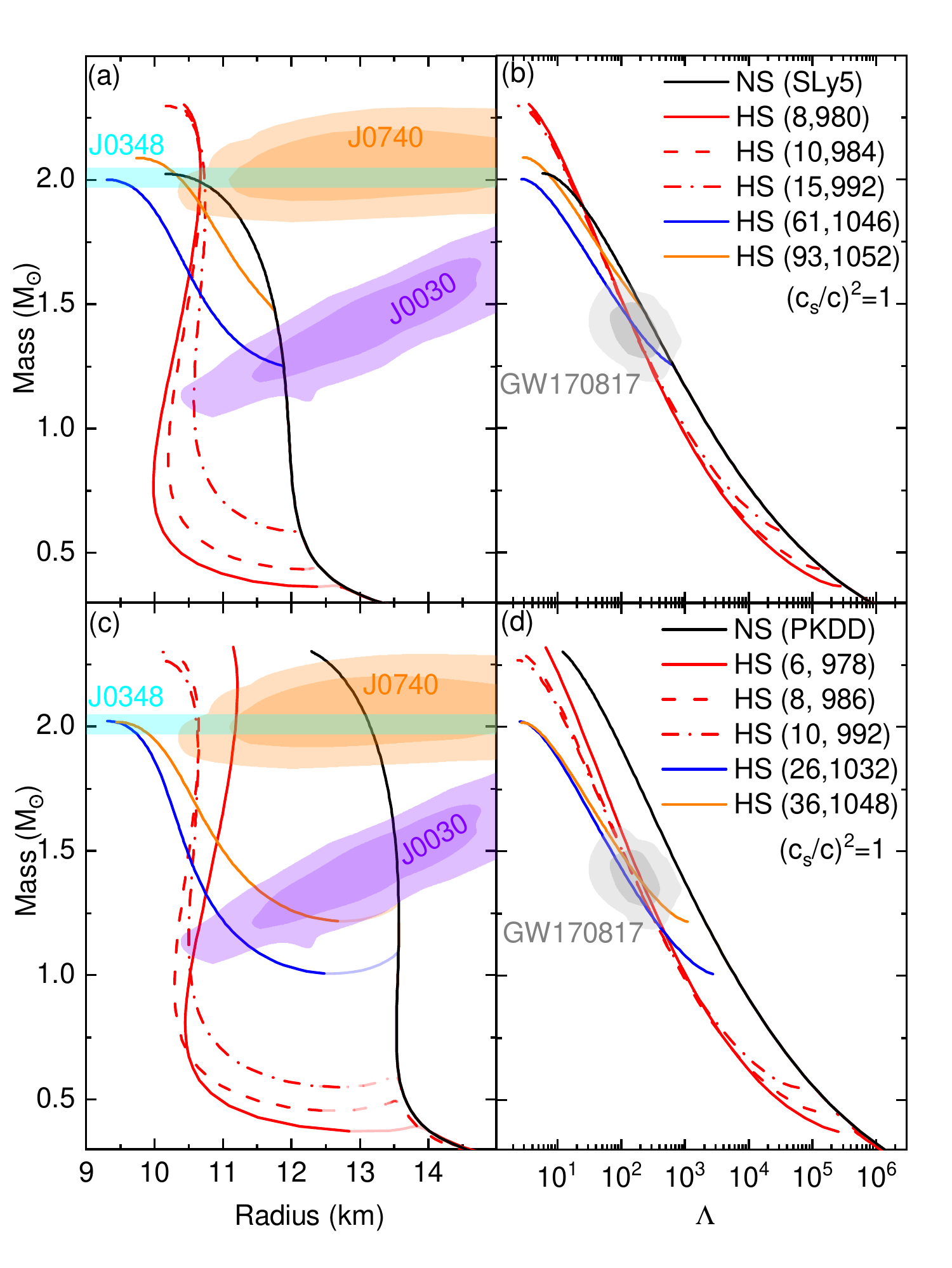}
\caption{Mass-radius (left) and mass-tidal deformability (right) representations for SLy5~\cite{Chabanat:1998} (top) and PKDD~\cite{Long:2004} (bottom). The observational contours are related to the analysis of: J0348+0432~\cite{Antoniadis:2013},  J0030+0451~\cite{Miller:2019}, J0740+6620~\cite{Miller:2021}, and GW170817~\cite{De:2018}. The lines represent predictions of a few selected EoS: purely nucleonic EoS for NS (black and solid lines), and the others (color lines) show examples of FOPT built upon nucleonic EoS considering the maximally stiff case ($c_s=c$) for the quark phase. The FOPT parameters $p_\fopt$ (MeV~fm$^{-3}$) and $\mu^*$ (MeV) are given in the inset for HS.}
\label{fig:MR}
\end{figure}

Over the last decade, the observation of CSs has entered into the era of precision measurements, which provides unprecedented possibilities for constraining the EoS. Among these constraints, the most impactful ones are shown in Fig.~\ref{fig:MR}: The detection of massive NSs by radio astronomers has pushed up the maximum mass limit, which is presently around 2M$_\odot$~\cite{Demorest:2010,Arzoumanian:2018,Antoniadis:2013,Cromartie:2019,Linares:2018,Ozel:2016}. As a typical example, the maximum mass limit from J0348+0432~\cite{Antoniadis:2013} is shown in Fig.~\ref{fig:MR}.
In addition to the masses, the radii of a few millisecond pulsars (PSR J0030+0451~\cite{Miller:2019,Riley:2019} and PSR-J0740+6620~\cite{Miller:2021,Riley:2021}) have recently been extracted from X-rays observations, using NICER observatory~\cite{Gendreau:2017}. 
Finally, the event GW170817~\cite{Abbott:2017,Abbott:2018,Abbott:2019}, produced by the merger of two CSs, allowed the first estimation of the effective tidal deformability $\tilde{\Lambda}$ for dense matter~\cite{Flanagan:2008,Hinderer:2008,Damour:2009}, which is shown in the right panels in Fig.~\ref{fig:MR}. 

The link between the observation of CSs and the strong interaction is performed through the equation of state (EoS), i.e., the pressure $p$ versus the energy density $\varepsilon$. By solving the Tolman-Oppenheimer-Volkoff (TOV) equations for spherical non-rotating and non-magnetized stellar objects~\cite{Tolman:1939,Oppenheimer:1939}, the EoS can be transformed into observational CS properties, e.g., a relation between masses and radii. For such kind of direct comparison, it is important to accurately estimate systematical uncertainties. This can be performed for instance by solving the TOV equations for a large set of EoSs compatible with the current knowledge. In this way, the model uncertainties in the EoS can be turned into a contour in the macroscopic properties, e.g., masses, radii and tidal deformabilities, shown in Fig.~\ref{fig:MR}. The tidal deformability is well correlated to the compactness ($\beta=M/R$) of the compact star~\cite{Abbott:2017}, and together with the measure of the mass extracted from the gravitational waveform, it allows to infer the value of the radius~\cite{Tews:2019,Abbott:2019}. In some analyses, it is the radius prediction inferred from the tidal deformability, assuming agnostic EoS modeling or universal relations, which is directly used to select EoS models~\cite{Xie:2021}. 

A few examples of NS and HS EoSs are shown in Fig.~\ref{fig:MR}, comparing predictions based on the nucleonic SLy5~\cite{Chabanat:1998} (top panel) and PKDD~\cite{Long:2004} (bottom panel) EoSs. We employ in this study these two typical nucleonic EoS which differ by the density dependence of the symmetry energy: we consider a soft nucleonic model, represented by SLy5~\cite{Chabanat:1998} nuclear model and a stiff nucleonic model, represented by PKDD~\cite{Long:2004} relativistic Lagrangian. The choice of reducing the nuclear EoS to only two typical ones is also performed by other authors, see for instance Ref.~\cite{Han:2019}. The SLy5 model is compatible with recent chiral EFT predictions with a low value of the slope of the symmetry energy ($L_\textrm{sym}=48.3$~MeV)~\cite{Drischler:2019,Somasundaram:2021}, as well as with the analysis of the PREX-II experiment including also binding energies, charge radii and dipole polarisabilities in a set of finite nuclei~\cite{Reinhard:2021}. The stiff PKDD model ($L_\textrm{sym}=79.5$~MeV) is compatible with another analysis of PREX-II experiment~\cite{Reed:2021}. 

Note that all the EoSs considered here satisfy the experimental constraints of nuclear physics, and the observations of NSs such as $2$M$_\odot$ limit, the stability and causality conditions. 
Twin stars, i.e., two stars with same mass but different radii, may exist~\cite{Li:2022} but, in the present study, they are not considered since we impose a one-to-one correspondence between masses and radii.

By showing models and observational contours, Fig.~\ref{fig:MR} illustrates that it is difficult to find equations of state which reproduce equally well the contours from NICER observatory (left panels) and from GW170817 (right panel). For instance PKDD is out of the GW170817 contour but it is in very good agreement with the two other contours from NICER observatory. The opposite is observed for the other models which reproduce well GW170817. This tension has already been observed in previous studies, see for instance Refs.~\cite{Capano:2019eae,Guven:2020,Pang:2021,Dinh:2021}. 
The mass asymmetry of the binary CS predicted for GW170817 is explored in the following analysis but we do not confirm, as suggested in Refs.~\cite{Pankow:2018,Horvath:2019,Ferdman:2020}, that it could resolve the mismatch illustrated in Fig.1.
Because of this tension in the data, we adopt the common strategy: we employ GW170817 to select EoSs, by comparing different kind of binary systems: binary neutron stars (BNS), binary hybrid stars (BHS) and neutron star hybrid star (NSHS). This allows to determine whether one of these system is favored, in a Bayesian framework, by the present data. In a second step, we combine the prediction of these modelings together with NICER contour to infer the radius $R_{1.4}$ of a $1.4$M$_\odot$ compact star.

In order to describe hybrid stars, a first order phase transition is built on top of nucleonic EoSs. The quark phase is described by the constant sound speed approach inspired from the MIT bag model~\cite{Chodos:1974}. It requires three quantities: the pressure $p_\fopt$ at the entrance of the phase transition, the shift in energy density, reflecting the latent heat $\Delta\varepsilon_\fopt$, and finally, the sound speed $c_s$ in quark matter assumed to be constant. We have~\cite{Zdunik:2013,Alford:2013,Chamel:2013}:
\begin{equation}
\label{e1}
\varepsilon(p) =\begin{cases}\varepsilon_{\NM}(p) & p < p_\fopt \\\varepsilon_{\NM}(p_\fopt)+\Delta\varepsilon_\fopt+(p-p_\fopt)/\alpha & p \geq p_\fopt\end{cases}
\end{equation}
where $\varepsilon_{\NM}$ is the nucleonic matter energy density, and $\alpha$ is related to the sound speed as $\alpha=(c_{s}/c)^{2}$, where $c$ is the speed of light in the vacuum. In the following, we use units where $c=1$. A similar approach has recently been employed in~\cite{Han:2019,Xie:2021,Somasundaram:2022a,Somasundaram:2022b}, leading to the possible existence of a third branch of compact stars~\cite{Alford:2017}. In our study, we use these modeling to explore the nature of CS favored by the present data.

\begin{figure}[t]
\centering
\includegraphics[width=0.48\textwidth]{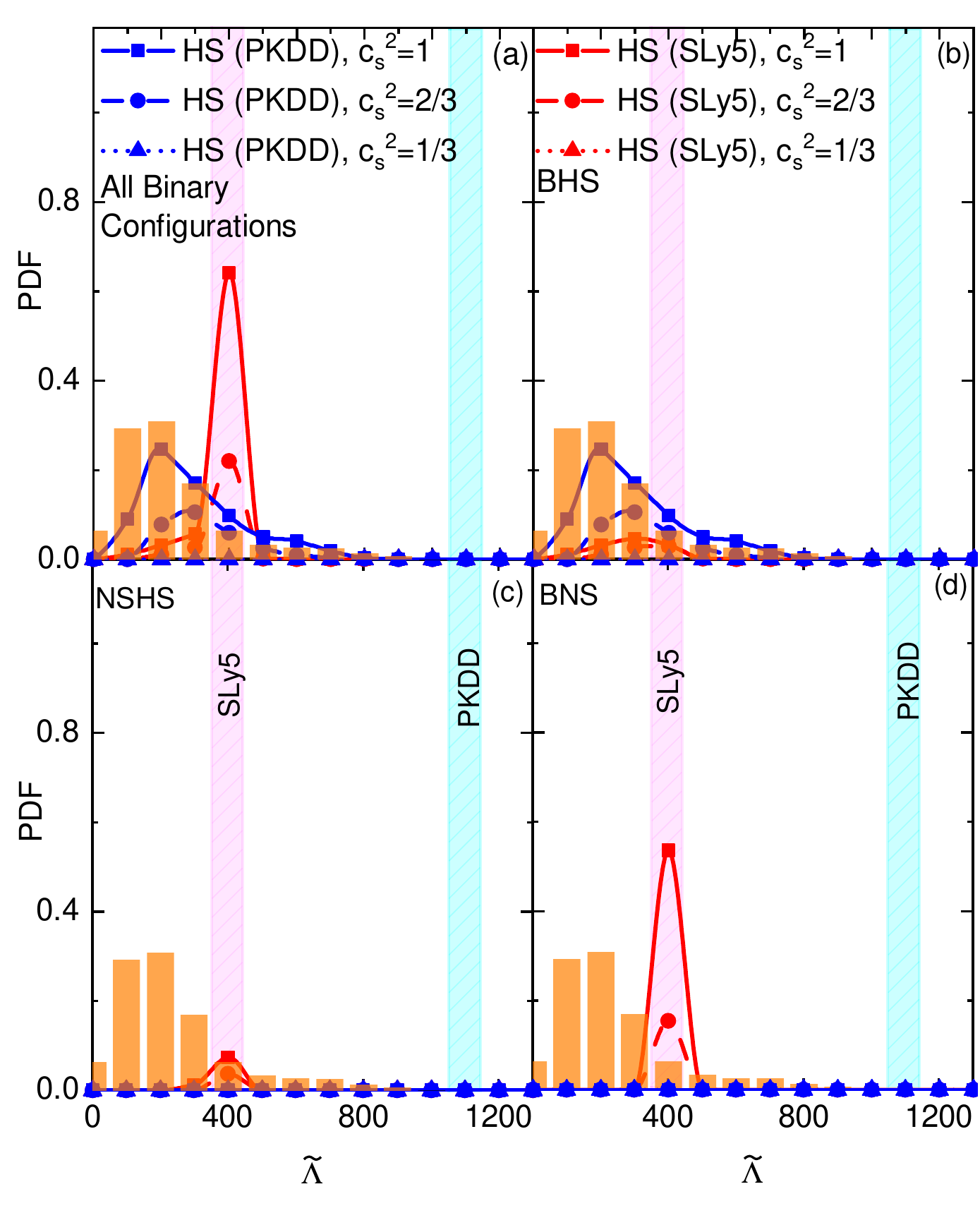}
\caption{Posterior PDF function of $\tilde{\Lambda}$ obtained for BNS, BHS and NSHS configurations, compared to the data from Ref.~\cite{De:2018}. The different line styles correspond to different sound speeds $c_s^2=1/3$, $2/3$ and $1$. The two vertical bands represent the tidal deformabilities obtained for the purely nucleonic EoSs, see the panel associated to binary configurations.}
\label{fig:Lambda}
\end{figure}

\begin{figure}[t]
\centering
\includegraphics[width=0.48\textwidth]{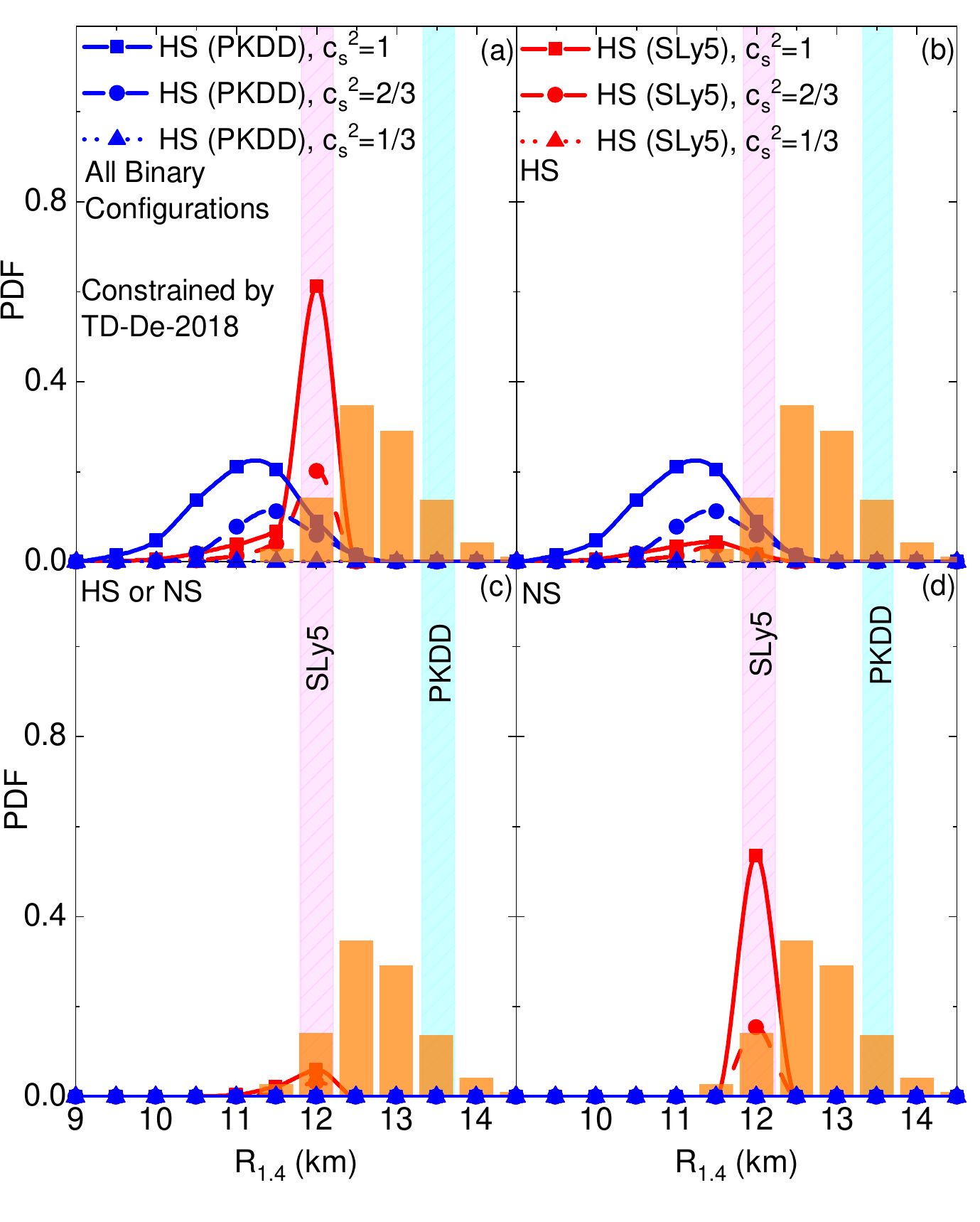}
\caption{Same as Fig.~\ref{fig:Lambda} but as function of the radius $R_{1.4}$. The data shown in the graphs are deduced from the analysis of PSR J0030+0451 by Miller \textsl{et al.}~\cite{Miller:2019} where the PDF is sliced for masses between 1.35 and 1.45M$_\odot$. The two vertical bands represent the radii obtained from the purely nucleonic EoSs, see the panel associated to binary configurations.}
\label{fig:radius}
\end{figure}

In practice we use the parameters $p_\fopt$ and $\mu^*$, which represents the chemical potential of quark matter at zero pressure, and from which the latent heat $\Delta\varepsilon_\fopt$ can be recovered as 
\begin{equation}
\Delta\varepsilon_\fopt = \frac{p_\fopt}{ \alpha }  \left[ \frac{1 +  \alpha }{ 
\left[\frac{ \mu_{\NM}(p_\fopt)}{\mu^*}\right]^{ \frac{1+\alpha}{\alpha}} -1} +1    \right] -  \varepsilon_{\NM}(p_\fopt) \, .
\end{equation}
We consider the following ranges for the model parameters and we consider uniform priors: $\mu^*=925\pm 75$~MeV, suggested from color-superconducting quark matter calculations~\cite{Agrawal:2010}; the sound speed parameter $\alpha$ is fixed to three possible values: $\alpha=1/3$ (conformal limit), $2/3$ and $1$ (causal limit); and finally, $p_\fopt$ is an unconstrained parameter with unknown boundaries. Here, we vary $p_\fopt$ from 6 to 500~MeV~fm$^{-3}$, guided by the vanishing of the posterior distribution.

It should be noted that in the present study, quark matter may appear already around (and above) saturation density, since we explore a large space for the sound speed, ranging between $1/\sqrt{3}$ and 1. While other analyses also explore a large space (see for instance Ref.~\cite{Xie:2021}) which predict quark matter to appear on the average for densities around 1.5$n_\sat$, some analyses of quark matter predict its appearance above $3n_\sat$~\cite{Pfaff:2022,Ayriyan:2019} or above $1.6$M$_\odot$~\cite{Parisi:2021}. These predictions may be related to the choice of the sound speed prior, which is limited to low values in the quark phase in these analyses while we explore a wider domain in the present study.

We employ the Bayesian approach to compare our model predictions, represented by a set of EoS parameters $\{a_i\}$ with the present data~\cite{Steiner:2015}. The probability associated to a given model considering a set of data, the so-called the posterior probability, is 
\begin{equation}\label{e13}
P(\{a_i\} \vert \textrm{data})\sim P(\textrm{data}\vert \{a_i\})\times P(\{a_i\}),
\end{equation}
where $P(\textrm{data}\vert \{a_i\})$ is the likelihood function representing the ability of the model to reproduce a set of measurements and $P(\{a_i\})$ is the prior probability, which represents the \textsl{a-priori} knowledge on the model parameters. We vary the parameters controlling the FOPT, $\alpha$, $p_\fopt$ and $\mu^*$, which are yet unconstrained parameters. The likelihood probability entering Eq.~\eqref{e13} is defined as,
\begin{equation}\label{e14}
P(\mathrm{data}\vert \{a_i\})=w_\mathrm{filter}(\{a_i\}) \times p_{\tilde{\Lambda}}  \, ,
\end{equation}
where $w_\mathrm{filter}(\{ a_i\})$ is a pass-band type filter which selects only the viable models with maximum mass larger than 2M$_\odot$ and $p_{\tilde{\Lambda}}$ expresses the ability of the model to reproduce the observed probability density function (PDF) for $\tilde{\Lambda}$ deduced from the GW signal, considering the wave-form analysis from Ref.~\cite{De:2018} and Ref.~\cite{Abbott:2019}. To do so, the effective tidal deformability $\tilde{\Lambda}(q)$ is averaged over the mass ratio $q$ of the binary system in the range $[0.73,1]$, see Ref.~\cite{Guven:2020} for more details. 

\begin{table}[t]
\tabcolsep=0.25cm
\def\arraystretch{1.4}
\begin{tabular}{cccccc}
\hline\hline
& EOS & $ c_s^2$ &  BHS & NSHS & BNS\\
\hline
\parbox[t]{2mm}{\multirow{3}{*}{\rotatebox[origin=c]{90}{soft}}} & SLy5 & 1/3 & $0.00/0.00$ & $0.00/0.00$ & $0.00/0.00$ \\
& SLy5 & 2/3 & $0.07/0.07$ & $0.04/0.04$ & $0.07/0.07$ \\
& SLy5 & 1   & $0.12/0.15$ & $0.08/0.08$ & $0.06/0.07$ \\
\hline
\parbox[t]{2mm}{\multirow{3}{*}{\rotatebox[origin=c]{90}{stiff}}}& PKDD & 1/3 & $0.00/0.00$ & $0.00/0.00$ & $0.00/0.00$ \\
& PKDD & 2/3 & $\mathbf{0.28/0.27}$ & $0.00/0.00$ & $0.00/0.00$ \\
& PKDD & 1   & $\mathbf{0.65/0.65}$ & $0.00/0.00$ & $0.00/0.00$ \\
\hline\hline        
\end{tabular}
\caption{Overlap between the model prediction and the PDF associated to GW170817 (from Ref.~\cite{De:2018}/\cite{Abbott:2019}), as function of the nucleonic EoS, the sound speed in quark matter and the nature of the binary system. The largest overlaps are stressed in bold.}
\label{tab:overlap}
\end{table}

\begin{table}[t]
\tabcolsep=0.1cm
\def\arraystretch{1.4}
\begin{tabular}{ccccc}
\hline\hline
& EOS & Configuration & $ c_s^2$ & $R_{1.4}$ Radius \\
 & & & &km \\
\hline
\parbox[t]{2mm}{\multirow{7}{*}{\rotatebox[origin=c]{90}{soft}}} & SLy5 & NS & 2/3 & $12.00 (20)/12.00(20)$ \\
& SLy5 & NS & 1   & $12.00(20)/12.00(20)$ \\
& SLy5 & HS & 2/3 & $11.86(31)/11.97(37)$ \\
& SLy5 & HS & 1   & $11.84(32)11.95(38)$ \\
& SLy5 & HS or NS & 2/3 & $11.98(22)11.98(22)$ \\
& SLy5 & HS or NS & 1   & $11.96(24)/11.97(23)$ \\
& Average & & & $11.94(31)/11.98(32)$ \\
\hline
\parbox[t]{2mm}{\multirow{3}{*}{\rotatebox[origin=c]{90}{stiff}}} & PKDD & HS & 2/3 & $12.06(40)/12.35(37)$ \\
& PKDD & HS & 1   & $11.98 (39)/12.27(41)$\\
& Average & & & $12.02(39)/12.31(40)$ \\
& Global Average & & &
$12.22(45)$ \\
\hline\hline
\end{tabular}
\caption{Predictions for the $R_{1.4}$ radius constrained by both GW170817 and NICER observation of PSR J0030+0451, as function of the nature of the CS.}
\label{tab:radius}
\end{table}

The posterior probability as a function of the effective tidal deformability $\tilde{\Lambda}$, is shown in Fig.~\ref{fig:Lambda} for BNS, BHS and NSHS configurations considering the parameter estimation from Ref.~\cite{De:2018}. See the supplementary material for a similar figure where parameter estimation from Ref.~\cite{Abbott:2019} is used instead.
For BNS systems, only soft nucleonic EoSs overlap the data since stiff nucleonic models predict too large tidal deformabilities. NSHS systems are not favored by either soft or stiff nucleonic EoS. Finally, BHS systems are preferred by stiff nucleonic models, overlapping at best the data, as shown in Tab.~\ref{tab:overlap} where bold values mark the best overlaps. Remark that there is a non-negligible overlap of the data for BNS systems with iso-soft EoS such as SLy5.
It can be remarked that, in all cases, the overlap is generally optimal for large sound speed, $c_s^2\gtrsim 2/3$, implying a stiff quark matter phase.
Considering Fig.~\ref{fig:Lambda} and Tab.~\ref{tab:overlap}, we conclude that current GW and radio-astronomy observations favors stiff nucleonic EoS compatible with nuclear physics and a BHS configuration for GW170817 event.
BNS configuration with soft nucleonic EoS cannot however be fully excluded.

The same EoSs are employed to predict $R_{1.4}$, as shown in 
Fig.~\ref{fig:radius}. Independently of their ability to describe GW170817, the two nucleonic EoS compatible with nuclear physics have a good overlap with the NICER data for NSs: they predict radii from $12.00$~km (SLy5) to $13.50$ (PKDD). Tab.~\ref{tab:radius} subsumes the $R_{1.4}$ predictions conditioned by GW170817, radio-astronomy and nuclear physics. We consider only the cases for which non-zero values are obtained in Tab.~\ref{tab:overlap}. The centroids are weakly dependent on the nucleonic modelling, going from about 11.84 to about 12.35~km. We obtain however that the stiff nucleonic EoS, favoring the HS configuration, predicts a radius $R_{1.4}=12.22(45)$~km, shown in Tab.~\ref{tab:radius}. The soft nucleonic EoS, which is not favored in Tab.~\ref{tab:overlap}, predicts a radius $R_{1.4}=12.00(20)$~km. These two predictions are quite close and considering moreover their uncertainties, they are not significantly different. 
Let us note that the uncertainties which we report are obtained for a given nucleonic EoS. We expect, in a future work, to obtain larger uncertainties by exploring more nucleonic EoS and therefore getting closer to the ones obtained in Refs~\cite{Dietrich:2020,Al-Mamun:2021}.
We then find that the radius of a $1.4$M$_\odot$ CS is weakly impacted by the nucleonic EoS as well as by the nature of CS within the two nucleonic scenarios that we have investigated. By weakly, we mean that the uncertainties originating from the nucleonic EoS are not the dominant ones. 

In conclusion, we have obtained that the cross-constraints from gravitational wave, radio-astronomy, x-ray observations and nuclear physics favor binary compact star systems with stiff nucleonic EoS complemented with stiff quark matter. The neutron star-hybrid star configuration is not statistically favored for GW170817 event. The present analysis supports the existence of $1.4$~M$_\odot$ mass hybrid star, with a low value for the phase transition ($n_\mathrm{tr}\in[0.18,0.35]~\mathrm{fm}^{-3}$).
Our analysis illustrates the complementary between nuclear physics and astrophysics for the understanding of dense matter in CS. Future tight constraints in the slope of the symmetry energy $L_\sym$ will further constrain the nature of CSs in GW170817, as well as in the future astrophysical observations, such as for instance binary CS mergers, radio-astronomy or x-ray emission from millisecond pulsars.

This work is supported by the Scientific and Technological Research Council of Turkey (T\"{U}B\.{I}TAK) under project number MFAG-122F121 and the Yildiz Technical University under project number  FBA-2021-4229. J.M. and E. K. are both supported by the CNRS/IN2P3 MAC project.

\bibliography{biblio}

\begin{thebibliography}{54}%
\makeatletter
\providecommand \@ifxundefined [1]{%
 \@ifx{#1\undefined}
}%
\providecommand \@ifnum [1]{%
 \ifnum #1\expandafter \@firstoftwo
 \else \expandafter \@secondoftwo
 \fi
}%
\providecommand \@ifx [1]{%
 \ifx #1\expandafter \@firstoftwo
 \else \expandafter \@secondoftwo
 \fi
}%
\providecommand \natexlab [1]{#1}%
\providecommand \enquote  [1]{``#1''}%
\providecommand \bibnamefont  [1]{#1}%
\providecommand \bibfnamefont [1]{#1}%
\providecommand \citenamefont [1]{#1}%
\providecommand \href@noop [0]{\@secondoftwo}%
\providecommand \href [0]{\begingroup \@sanitize@url \@href}%
\providecommand \@href[1]{\@@startlink{#1}\@@href}%
\providecommand \@@href[1]{\endgroup#1\@@endlink}%
\providecommand \@sanitize@url [0]{\catcode `\\12\catcode `\$12\catcode
  `\&12\catcode `\#12\catcode `\^12\catcode `\_12\catcode `\%12\relax}%
\providecommand \@@startlink[1]{}%
\providecommand \@@endlink[0]{}%
\providecommand \url  [0]{\begingroup\@sanitize@url \@url }%
\providecommand \@url [1]{\endgroup\@href {#1}{\urlprefix }}%
\providecommand \urlprefix  [0]{URL }%
\providecommand \Eprint [0]{\href }%
\providecommand \doibase [0]{https://doi.org/}%
\providecommand \selectlanguage [0]{\@gobble}%
\providecommand \bibinfo  [0]{\@secondoftwo}%
\providecommand \bibfield  [0]{\@secondoftwo}%
\providecommand \translation [1]{[#1]}%
\providecommand \BibitemOpen [0]{}%
\providecommand \bibitemStop [0]{}%
\providecommand \bibitemNoStop [0]{.\EOS\space}%
\providecommand \EOS [0]{\spacefactor3000\relax}%
\providecommand \BibitemShut  [1]{\csname bibitem#1\endcsname}%
\let\auto@bib@innerbib\@empty
\bibitem [{\citenamefont {Alford}\ \emph {et~al.}(2008)\citenamefont {Alford},
  \citenamefont {Schmitt}, \citenamefont {Rajagopal},\ and\ \citenamefont
  {Sch\"afer}}]{Alford:2008}%
  \BibitemOpen
  \bibfield  {author} {\bibinfo {author} {\bibfnamefont {M.~G.}\ \bibnamefont
  {Alford}}, \bibinfo {author} {\bibfnamefont {A.}~\bibnamefont {Schmitt}},
  \bibinfo {author} {\bibfnamefont {K.}~\bibnamefont {Rajagopal}},\ and\
  \bibinfo {author} {\bibfnamefont {T.}~\bibnamefont {Sch\"afer}},\ }\bibfield
  {title} {\bibinfo {title} {Color superconductivity in dense quark matter},\
  }\href {https://doi.org/10.1103/RevModPhys.80.1455} {\bibfield  {journal}
  {\bibinfo  {journal} {Rev. Mod. Phys.}\ }\textbf {\bibinfo {volume} {80}},\
  \bibinfo {pages} {1455} (\bibinfo {year} {2008})}\BibitemShut {NoStop}%
\bibitem [{\citenamefont {Anglani}\ \emph {et~al.}(2014)\citenamefont
  {Anglani}, \citenamefont {Casalbuoni}, \citenamefont {Ciminale},
  \citenamefont {Ippolito}, \citenamefont {Gatto}, \citenamefont {Mannarelli},\
  and\ \citenamefont {Ruggieri}}]{Anglani:2014}%
  \BibitemOpen
  \bibfield  {author} {\bibinfo {author} {\bibfnamefont {R.}~\bibnamefont
  {Anglani}}, \bibinfo {author} {\bibfnamefont {R.}~\bibnamefont {Casalbuoni}},
  \bibinfo {author} {\bibfnamefont {M.}~\bibnamefont {Ciminale}}, \bibinfo
  {author} {\bibfnamefont {N.}~\bibnamefont {Ippolito}}, \bibinfo {author}
  {\bibfnamefont {R.}~\bibnamefont {Gatto}}, \bibinfo {author} {\bibfnamefont
  {M.}~\bibnamefont {Mannarelli}},\ and\ \bibinfo {author} {\bibfnamefont
  {M.}~\bibnamefont {Ruggieri}},\ }\bibfield  {title} {\bibinfo {title}
  {Crystalline color superconductors},\ }\href
  {https://doi.org/10.1103/RevModPhys.86.509} {\bibfield  {journal} {\bibinfo
  {journal} {Rev. Mod. Phys.}\ }\textbf {\bibinfo {volume} {86}},\ \bibinfo
  {pages} {509} (\bibinfo {year} {2014})}\BibitemShut {NoStop}%
\bibitem [{\citenamefont {Collaboration}(2019)}]{HADES:2019}%
  \BibitemOpen
  \bibfield  {author} {\bibinfo {author} {\bibfnamefont {T.~H.}\ \bibnamefont
  {Collaboration}},\ }\bibfield  {title} {\bibinfo {title} {Probing dense
  baryon-rich matter with virtual photons},\ }\href@noop {} {\bibfield
  {journal} {\bibinfo  {journal} {Nat. Phys.}\ }\textbf {\bibinfo {volume}
  {15}},\ \bibinfo {pages} {1040} (\bibinfo {year} {2019})}\BibitemShut
  {NoStop}%
\bibitem [{\citenamefont {Chabanat}\ \emph {et~al.}(1998)\citenamefont
  {Chabanat}, \citenamefont {Bonche}, \citenamefont {Haensel}, \citenamefont
  {Meyer},\ and\ \citenamefont {Schaeffer}}]{Chabanat:1998}%
  \BibitemOpen
  \bibfield  {author} {\bibinfo {author} {\bibfnamefont {E.}~\bibnamefont
  {Chabanat}}, \bibinfo {author} {\bibfnamefont {P.}~\bibnamefont {Bonche}},
  \bibinfo {author} {\bibfnamefont {P.}~\bibnamefont {Haensel}}, \bibinfo
  {author} {\bibfnamefont {J.}~\bibnamefont {Meyer}},\ and\ \bibinfo {author}
  {\bibfnamefont {R.}~\bibnamefont {Schaeffer}},\ }\bibfield  {title} {\bibinfo
  {title} {{A Skyrme parametrization from subnuclear to neutron star densities.
  2. Nuclei far from stablities}},\ }\href
  {https://doi.org/10.1016/S0375-9474(98)00180-8} {\bibfield  {journal}
  {\bibinfo  {journal} {Nucl. Phys. A}\ }\textbf {\bibinfo {volume} {635}},\
  \bibinfo {pages} {231} (\bibinfo {year} {1998})},\ \bibinfo {note} {[Erratum:
  Nucl.Phys.A 643, 441--441 (1998)]}\BibitemShut {NoStop}%
\bibitem [{\citenamefont {Long}\ \emph {et~al.}(2004)\citenamefont {Long},
  \citenamefont {Meng}, \citenamefont {VanGiai},\ and\ \citenamefont
  {Zhou}}]{Long:2004}%
  \BibitemOpen
  \bibfield  {author} {\bibinfo {author} {\bibfnamefont {W.}~\bibnamefont
  {Long}}, \bibinfo {author} {\bibfnamefont {J.}~\bibnamefont {Meng}}, \bibinfo
  {author} {\bibfnamefont {N.}~\bibnamefont {VanGiai}},\ and\ \bibinfo {author}
  {\bibfnamefont {S.-G.}\ \bibnamefont {Zhou}},\ }\bibfield  {title} {\bibinfo
  {title} {New effective interactions in relativistic mean field theory with
  nonlinear terms and density-dependent meson-nucleon coupling},\ }\href
  {https://doi.org/10.1103/PhysRevC.69.034319} {\bibfield  {journal} {\bibinfo
  {journal} {Phys. Rev. C}\ }\textbf {\bibinfo {volume} {69}},\ \bibinfo
  {pages} {034319} (\bibinfo {year} {2004})}\BibitemShut {NoStop}%
\bibitem [{\citenamefont {Antoniadis}\ \emph {et~al.}(2013)\citenamefont
  {Antoniadis} \emph {et~al.}}]{Antoniadis:2013}%
  \BibitemOpen
  \bibfield  {author} {\bibinfo {author} {\bibfnamefont {J.}~\bibnamefont
  {Antoniadis}} \emph {et~al.},\ }\bibfield  {title} {\bibinfo {title} {{A
  Massive Pulsar in a Compact Relativistic Binary}},\ }\href
  {https://doi.org/10.1126/science.1233232} {\bibfield  {journal} {\bibinfo
  {journal} {Science}\ }\textbf {\bibinfo {volume} {340}},\ \bibinfo {pages}
  {6131} (\bibinfo {year} {2013})},\ \Eprint {https://arxiv.org/abs/1304.6875}
  {arXiv:1304.6875 [astro-ph.HE]} \BibitemShut {NoStop}%
\bibitem [{\citenamefont {Miller}\ \emph {et~al.}(2019)\citenamefont {Miller},
  \citenamefont {Lamb}, \citenamefont {Dittmann}, \citenamefont {Bogdanov},
  \citenamefont {Arzoumanian}, \citenamefont {Gendreau}, \citenamefont
  {Guillot}, \citenamefont {Harding}, \citenamefont {Ho}, \citenamefont
  {Lattimer}, \citenamefont {Ludlam}, \citenamefont {Mahmoodifar},
  \citenamefont {Morsink}, \citenamefont {Ray}, \citenamefont {Strohmayer},
  \citenamefont {Wood}, \citenamefont {Enoto}, \citenamefont {Foster},
  \citenamefont {Okajima}, \citenamefont {Prigozhin},\ and\ \citenamefont
  {Soong}}]{Miller:2019}%
  \BibitemOpen
  \bibfield  {author} {\bibinfo {author} {\bibfnamefont {M.~C.}\ \bibnamefont
  {Miller}}, \bibinfo {author} {\bibfnamefont {F.~K.}\ \bibnamefont {Lamb}},
  \bibinfo {author} {\bibfnamefont {A.~J.}\ \bibnamefont {Dittmann}}, \bibinfo
  {author} {\bibfnamefont {S.}~\bibnamefont {Bogdanov}}, \bibinfo {author}
  {\bibfnamefont {Z.}~\bibnamefont {Arzoumanian}}, \bibinfo {author}
  {\bibfnamefont {K.~C.}\ \bibnamefont {Gendreau}}, \bibinfo {author}
  {\bibfnamefont {S.}~\bibnamefont {Guillot}}, \bibinfo {author} {\bibfnamefont
  {A.~K.}\ \bibnamefont {Harding}}, \bibinfo {author} {\bibfnamefont
  {W.~C.~G.}\ \bibnamefont {Ho}}, \bibinfo {author} {\bibfnamefont {J.~M.}\
  \bibnamefont {Lattimer}}, \bibinfo {author} {\bibfnamefont {R.~M.}\
  \bibnamefont {Ludlam}}, \bibinfo {author} {\bibfnamefont {S.}~\bibnamefont
  {Mahmoodifar}}, \bibinfo {author} {\bibfnamefont {S.~M.}\ \bibnamefont
  {Morsink}}, \bibinfo {author} {\bibfnamefont {P.~S.}\ \bibnamefont {Ray}},
  \bibinfo {author} {\bibfnamefont {T.~E.}\ \bibnamefont {Strohmayer}},
  \bibinfo {author} {\bibfnamefont {K.~S.}\ \bibnamefont {Wood}}, \bibinfo
  {author} {\bibfnamefont {T.}~\bibnamefont {Enoto}}, \bibinfo {author}
  {\bibfnamefont {R.}~\bibnamefont {Foster}}, \bibinfo {author} {\bibfnamefont
  {T.}~\bibnamefont {Okajima}}, \bibinfo {author} {\bibfnamefont
  {G.}~\bibnamefont {Prigozhin}},\ and\ \bibinfo {author} {\bibfnamefont
  {Y.}~\bibnamefont {Soong}},\ }\bibfield  {title} {\bibinfo {title} {Psr
  j0030+0451 mass and radius from nicer data and implications for the
  properties of neutron star matter},\ }\href
  {https://doi.org/10.3847/2041-8213/ab50c5} {\bibfield  {journal} {\bibinfo
  {journal} {The Astrophysical Journal Letters}\ }\textbf {\bibinfo {volume}
  {887}},\ \bibinfo {pages} {L24} (\bibinfo {year} {2019})}\BibitemShut
  {NoStop}%
\bibitem [{\citenamefont {Miller}\ \emph {et~al.}(2021)\citenamefont {Miller},
  \citenamefont {Lamb}, \citenamefont {Dittmann}, \citenamefont {Bogdanov},
  \citenamefont {Arzoumanian}, \citenamefont {Gendreau}, \citenamefont
  {Guillot}, \citenamefont {Ho}, \citenamefont {Lattimer}, \citenamefont
  {Loewenstein}, \citenamefont {Morsink}, \citenamefont {Ray}, \citenamefont
  {Wolff}, \citenamefont {Baker}, \citenamefont {Cazeau}, \citenamefont
  {Manthripragada}, \citenamefont {Markwardt}, \citenamefont {Okajima},
  \citenamefont {Pollard}, \citenamefont {Cognard}, \citenamefont {Cromartie},
  \citenamefont {Fonseca}, \citenamefont {Guillemot}, \citenamefont {Kerr},
  \citenamefont {Parthasarathy}, \citenamefont {Pennucci}, \citenamefont
  {Ransom},\ and\ \citenamefont {Stairs}}]{Miller:2021}%
  \BibitemOpen
  \bibfield  {author} {\bibinfo {author} {\bibfnamefont {M.~C.}\ \bibnamefont
  {Miller}}, \bibinfo {author} {\bibfnamefont {F.~K.}\ \bibnamefont {Lamb}},
  \bibinfo {author} {\bibfnamefont {A.~J.}\ \bibnamefont {Dittmann}}, \bibinfo
  {author} {\bibfnamefont {S.}~\bibnamefont {Bogdanov}}, \bibinfo {author}
  {\bibfnamefont {Z.}~\bibnamefont {Arzoumanian}}, \bibinfo {author}
  {\bibfnamefont {K.~C.}\ \bibnamefont {Gendreau}}, \bibinfo {author}
  {\bibfnamefont {S.}~\bibnamefont {Guillot}}, \bibinfo {author} {\bibfnamefont
  {W.~C.~G.}\ \bibnamefont {Ho}}, \bibinfo {author} {\bibfnamefont {J.~M.}\
  \bibnamefont {Lattimer}}, \bibinfo {author} {\bibfnamefont {M.}~\bibnamefont
  {Loewenstein}}, \bibinfo {author} {\bibfnamefont {S.~M.}\ \bibnamefont
  {Morsink}}, \bibinfo {author} {\bibfnamefont {P.~S.}\ \bibnamefont {Ray}},
  \bibinfo {author} {\bibfnamefont {M.~T.}\ \bibnamefont {Wolff}}, \bibinfo
  {author} {\bibfnamefont {C.~L.}\ \bibnamefont {Baker}}, \bibinfo {author}
  {\bibfnamefont {T.}~\bibnamefont {Cazeau}}, \bibinfo {author} {\bibfnamefont
  {S.}~\bibnamefont {Manthripragada}}, \bibinfo {author} {\bibfnamefont
  {C.~B.}\ \bibnamefont {Markwardt}}, \bibinfo {author} {\bibfnamefont
  {T.}~\bibnamefont {Okajima}}, \bibinfo {author} {\bibfnamefont
  {S.}~\bibnamefont {Pollard}}, \bibinfo {author} {\bibfnamefont
  {I.}~\bibnamefont {Cognard}}, \bibinfo {author} {\bibfnamefont {H.~T.}\
  \bibnamefont {Cromartie}}, \bibinfo {author} {\bibfnamefont {E.}~\bibnamefont
  {Fonseca}}, \bibinfo {author} {\bibfnamefont {L.}~\bibnamefont {Guillemot}},
  \bibinfo {author} {\bibfnamefont {M.}~\bibnamefont {Kerr}}, \bibinfo {author}
  {\bibfnamefont {A.}~\bibnamefont {Parthasarathy}}, \bibinfo {author}
  {\bibfnamefont {T.~T.}\ \bibnamefont {Pennucci}}, \bibinfo {author}
  {\bibfnamefont {S.}~\bibnamefont {Ransom}},\ and\ \bibinfo {author}
  {\bibfnamefont {I.}~\bibnamefont {Stairs}},\ }\bibfield  {title} {\bibinfo
  {title} {The radius of psr j0740+6620 from nicer and xmm-newton data},\
  }\href {https://doi.org/10.3847/2041-8213/ac089b} {\bibfield  {journal}
  {\bibinfo  {journal} {The Astrophysical Journal Letters}\ }\textbf {\bibinfo
  {volume} {918}},\ \bibinfo {pages} {L28} (\bibinfo {year}
  {2021})}\BibitemShut {NoStop}%
\bibitem [{\citenamefont {De}\ \emph {et~al.}(2018)\citenamefont {De},
  \citenamefont {Finstad}, \citenamefont {Lattimer}, \citenamefont {Brown},
  \citenamefont {Berger},\ and\ \citenamefont {Biwer}}]{De:2018}%
  \BibitemOpen
  \bibfield  {author} {\bibinfo {author} {\bibfnamefont {S.}~\bibnamefont
  {De}}, \bibinfo {author} {\bibfnamefont {D.}~\bibnamefont {Finstad}},
  \bibinfo {author} {\bibfnamefont {J.~M.}\ \bibnamefont {Lattimer}}, \bibinfo
  {author} {\bibfnamefont {D.~A.}\ \bibnamefont {Brown}}, \bibinfo {author}
  {\bibfnamefont {E.}~\bibnamefont {Berger}},\ and\ \bibinfo {author}
  {\bibfnamefont {C.~M.}\ \bibnamefont {Biwer}},\ }\bibfield  {title} {\bibinfo
  {title} {Tidal deformabilities and radii of neutron stars from the
  observation of gw170817},\ }\href
  {https://doi.org/10.1103/PhysRevLett.121.091102} {\bibfield  {journal}
  {\bibinfo  {journal} {Phys. Rev. Lett.}\ }\textbf {\bibinfo {volume} {121}},\
  \bibinfo {pages} {091102} (\bibinfo {year} {2018})}\BibitemShut {NoStop}%
\bibitem [{\citenamefont {Demorest}\ \emph {et~al.}(2010)\citenamefont
  {Demorest}, \citenamefont {Pennucci}, \citenamefont {Ransom}, \citenamefont
  {Roberts},\ and\ \citenamefont {Hessels}}]{Demorest:2010}%
  \BibitemOpen
  \bibfield  {author} {\bibinfo {author} {\bibfnamefont {P.}~\bibnamefont
  {Demorest}}, \bibinfo {author} {\bibfnamefont {T.}~\bibnamefont {Pennucci}},
  \bibinfo {author} {\bibfnamefont {S.}~\bibnamefont {Ransom}}, \bibinfo
  {author} {\bibfnamefont {M.}~\bibnamefont {Roberts}},\ and\ \bibinfo {author}
  {\bibfnamefont {J.}~\bibnamefont {Hessels}},\ }\bibfield  {title} {\bibinfo
  {title} {{Shapiro Delay Measurement of A Two Solar Mass Neutron Star}},\
  }\href {https://doi.org/10.1038/nature09466} {\bibfield  {journal} {\bibinfo
  {journal} {Nature}\ }\textbf {\bibinfo {volume} {467}},\ \bibinfo {pages}
  {1081} (\bibinfo {year} {2010})},\ \Eprint {https://arxiv.org/abs/1010.5788}
  {arXiv:1010.5788 [astro-ph.HE]} \BibitemShut {NoStop}%
\bibitem [{\citenamefont {Arzoumanian}\ \emph {et~al.}(2018)\citenamefont
  {Arzoumanian} \emph {et~al.}}]{Arzoumanian:2018}%
  \BibitemOpen
  \bibfield  {author} {\bibinfo {author} {\bibfnamefont {Z.}~\bibnamefont
  {Arzoumanian}} \emph {et~al.} (\bibinfo {collaboration} {NANOGrav}),\
  }\bibfield  {title} {\bibinfo {title} {{The NANOGrav 11-year Data Set:
  High-precision timing of 45 Millisecond Pulsars}},\ }\href
  {https://doi.org/10.3847/1538-4365/aab5b0} {\bibfield  {journal} {\bibinfo
  {journal} {Astrophys. J. Suppl.}\ }\textbf {\bibinfo {volume} {235}},\
  \bibinfo {pages} {37} (\bibinfo {year} {2018})},\ \Eprint
  {https://arxiv.org/abs/1801.01837} {arXiv:1801.01837 [astro-ph.HE]}
  \BibitemShut {NoStop}%
\bibitem [{\citenamefont {Cromartie}\ \emph {et~al.}(2019)\citenamefont
  {Cromartie} \emph {et~al.}}]{Cromartie:2019}%
  \BibitemOpen
  \bibfield  {author} {\bibinfo {author} {\bibfnamefont {H.~T.}\ \bibnamefont
  {Cromartie}} \emph {et~al.},\ }\bibfield  {title} {\bibinfo {title}
  {{Relativistic Shapiro delay measurements of an extremely massive millisecond
  pulsar}},\ }\href {https://doi.org/10.1038/s41550-019-0880-2} {\bibfield
  {journal} {\bibinfo  {journal} {Nature Astron.}\ }\textbf {\bibinfo {volume}
  {4}},\ \bibinfo {pages} {72} (\bibinfo {year} {2019})},\ \Eprint
  {https://arxiv.org/abs/1904.06759} {arXiv:1904.06759 [astro-ph.HE]}
  \BibitemShut {NoStop}%
\bibitem [{\citenamefont {Linares}\ \emph {et~al.}(2018)\citenamefont
  {Linares}, \citenamefont {Shahbaz},\ and\ \citenamefont
  {Casares}}]{Linares:2018}%
  \BibitemOpen
  \bibfield  {author} {\bibinfo {author} {\bibfnamefont {M.}~\bibnamefont
  {Linares}}, \bibinfo {author} {\bibfnamefont {T.}~\bibnamefont {Shahbaz}},\
  and\ \bibinfo {author} {\bibfnamefont {J.}~\bibnamefont {Casares}},\
  }\bibfield  {title} {\bibinfo {title} {{Peering into the dark side: Magnesium
  lines establish a massive neutron star in PSR J2215+5135}},\ }\href
  {https://doi.org/10.3847/1538-4357/aabde6} {\bibfield  {journal} {\bibinfo
  {journal} {Astrophys. J.}\ }\textbf {\bibinfo {volume} {859}},\ \bibinfo
  {pages} {54} (\bibinfo {year} {2018})},\ \Eprint
  {https://arxiv.org/abs/1805.08799} {arXiv:1805.08799 [astro-ph.HE]}
  \BibitemShut {NoStop}%
\bibitem [{\citenamefont {\"{O}zel}\ and\ \citenamefont
  {Freire}(2016)}]{Ozel:2016}%
  \BibitemOpen
  \bibfield  {author} {\bibinfo {author} {\bibfnamefont {F.}~\bibnamefont
  {\"{O}zel}}\ and\ \bibinfo {author} {\bibfnamefont {P.}~\bibnamefont
  {Freire}},\ }\bibfield  {title} {\bibinfo {title} {Masses, radii, and the
  equation of state of neutron stars},\ }\href
  {https://doi.org/10.1146/annurev-astro-081915-023322} {\bibfield  {journal}
  {\bibinfo  {journal} {Annual Review of Astronomy and Astrophysics}\ }\textbf
  {\bibinfo {volume} {54}},\ \bibinfo {pages} {401} (\bibinfo {year} {2016})},\
  \Eprint
  {https://arxiv.org/abs/https://doi.org/10.1146/annurev-astro-081915-023322}
  {https://doi.org/10.1146/annurev-astro-081915-023322} \BibitemShut {NoStop}%
\bibitem [{\citenamefont {Riley}\ \emph {et~al.}(2019)\citenamefont {Riley}
  \emph {et~al.}}]{Riley:2019}%
  \BibitemOpen
  \bibfield  {author} {\bibinfo {author} {\bibfnamefont {T.~E.}\ \bibnamefont
  {Riley}} \emph {et~al.},\ }\bibfield  {title} {\bibinfo {title} {{A $NICER$
  View of PSR J0030+0451: Millisecond Pulsar Parameter Estimation}},\ }\href
  {https://doi.org/10.3847/2041-8213/ab481c} {\bibfield  {journal} {\bibinfo
  {journal} {Astrophys. J. Lett.}\ }\textbf {\bibinfo {volume} {887}},\
  \bibinfo {pages} {L21} (\bibinfo {year} {2019})},\ \Eprint
  {https://arxiv.org/abs/1912.05702} {arXiv:1912.05702 [astro-ph.HE]}
  \BibitemShut {NoStop}%
\bibitem [{\citenamefont {Riley}\ \emph {et~al.}(2021)\citenamefont {Riley}
  \emph {et~al.}}]{Riley:2021}%
  \BibitemOpen
  \bibfield  {author} {\bibinfo {author} {\bibfnamefont {T.~E.}\ \bibnamefont
  {Riley}} \emph {et~al.},\ }\bibfield  {title} {\bibinfo {title} {{A NICER
  View of the Massive Pulsar PSR J0740+6620 Informed by Radio Timing and
  XMM-Newton Spectroscopy}},\ }\href {https://doi.org/10.3847/2041-8213/ac0a81}
  {\bibfield  {journal} {\bibinfo  {journal} {Astrophys. J. Lett.}\ }\textbf
  {\bibinfo {volume} {918}},\ \bibinfo {pages} {L27} (\bibinfo {year}
  {2021})},\ \Eprint {https://arxiv.org/abs/2105.06980} {arXiv:2105.06980
  [astro-ph.HE]} \BibitemShut {NoStop}%
\bibitem [{\citenamefont {{Gendreau}}\ and\ \citenamefont
  {{Arzoumanian}}(2017)}]{Gendreau:2017}%
  \BibitemOpen
  \bibfield  {author} {\bibinfo {author} {\bibfnamefont {K.}~\bibnamefont
  {{Gendreau}}}\ and\ \bibinfo {author} {\bibfnamefont {Z.}~\bibnamefont
  {{Arzoumanian}}},\ }\bibfield  {title} {\bibinfo {title} {{Searching for a
  pulse}},\ }\href {https://doi.org/10.1038/s41550-017-0301-3} {\bibfield
  {journal} {\bibinfo  {journal} {Nature Astronomy}\ }\textbf {\bibinfo
  {volume} {1}},\ \bibinfo {pages} {895} (\bibinfo {year} {2017})}\BibitemShut
  {NoStop}%
\bibitem [{\citenamefont {Abbott}\ \emph {et~al.}(2017)\citenamefont {Abbott},
  \citenamefont {Abbott}, \citenamefont {Abbott} \emph {et~al.}}]{Abbott:2017}%
  \BibitemOpen
  \bibfield  {author} {\bibinfo {author} {\bibfnamefont {B.~P.}\ \bibnamefont
  {Abbott}}, \bibinfo {author} {\bibfnamefont {R.}~\bibnamefont {Abbott}},
  \bibinfo {author} {\bibfnamefont {T.~D.}\ \bibnamefont {Abbott}}, \emph
  {et~al.} (\bibinfo {collaboration} {LIGO Scientific Collaboration and Virgo
  Collaboration}),\ }\bibfield  {title} {\bibinfo {title} {{GW170817:
  Observation of Gravitational Waves from a Binary Neutron Star Inspiral}},\
  }\href {https://doi.org/10.1103/PhysRevLett.119.161101} {\bibfield  {journal}
  {\bibinfo  {journal} {\prl}\ }\textbf {\bibinfo {volume} {119}},\ \bibinfo
  {eid} {161101} (\bibinfo {year} {2017})},\ \Eprint
  {https://arxiv.org/abs/1710.05832} {arXiv:1710.05832 [gr-qc]} \BibitemShut
  {NoStop}%
\bibitem [{\citenamefont {Abbott}\ \emph {et~al.}(2018)\citenamefont {Abbott},
  \citenamefont {Abbott}, \citenamefont {Abbott} \emph {et~al.}}]{Abbott:2018}%
  \BibitemOpen
  \bibfield  {author} {\bibinfo {author} {\bibfnamefont {B.~P.}\ \bibnamefont
  {Abbott}}, \bibinfo {author} {\bibfnamefont {R.}~\bibnamefont {Abbott}},
  \bibinfo {author} {\bibfnamefont {T.~D.}\ \bibnamefont {Abbott}}, \emph
  {et~al.} (\bibinfo {collaboration} {LIGO Scientific Collaboration and Virgo
  Collaboration}),\ }\bibfield  {title} {\bibinfo {title} {{GW170817:
  Measurements of Neutron Star Radii and Equation of State}},\ }\href
  {https://doi.org/10.1103/PhysRevLett.121.161101} {\bibfield  {journal}
  {\bibinfo  {journal} {\prl}\ }\textbf {\bibinfo {volume} {121}},\ \bibinfo
  {eid} {161101} (\bibinfo {year} {2018})}\BibitemShut {NoStop}%
\bibitem [{\citenamefont {Abbott}\ \emph {et~al.}(2019)\citenamefont {Abbott},
  \citenamefont {Abbott}, \citenamefont {Abbott} \emph {et~al.}}]{Abbott:2019}%
  \BibitemOpen
  \bibfield  {author} {\bibinfo {author} {\bibfnamefont {B.~P.}\ \bibnamefont
  {Abbott}}, \bibinfo {author} {\bibfnamefont {R.}~\bibnamefont {Abbott}},
  \bibinfo {author} {\bibfnamefont {T.~D.}\ \bibnamefont {Abbott}}, \emph
  {et~al.} (\bibinfo {collaboration} {LIGO Scientific Collaboration and Virgo
  Collaboration}),\ }\bibfield  {title} {\bibinfo {title} {Properties of the
  binary neutron star merger gw170817},\ }\href
  {https://doi.org/10.1103/PhysRevX.9.011001} {\bibfield  {journal} {\bibinfo
  {journal} {Phys. Rev. X}\ }\textbf {\bibinfo {volume} {9}},\ \bibinfo {pages}
  {011001} (\bibinfo {year} {2019})}\BibitemShut {NoStop}%
\bibitem [{\citenamefont {Flanagan}\ and\ \citenamefont
  {Hinderer}(2008)}]{Flanagan:2008}%
  \BibitemOpen
  \bibfield  {author} {\bibinfo {author} {\bibfnamefont {E.~E.}\ \bibnamefont
  {Flanagan}}\ and\ \bibinfo {author} {\bibfnamefont {T.}~\bibnamefont
  {Hinderer}},\ }\bibfield  {title} {\bibinfo {title} {Constraining
  neutron-star tidal love numbers with gravitational-wave detectors},\ }\href
  {https://doi.org/10.1103/PhysRevD.77.021502} {\bibfield  {journal} {\bibinfo
  {journal} {Phys. Rev. D}\ }\textbf {\bibinfo {volume} {77}},\ \bibinfo
  {pages} {021502(R)} (\bibinfo {year} {2008})}\BibitemShut {NoStop}%
\bibitem [{\citenamefont {Hinderer}(2008)}]{Hinderer:2008}%
  \BibitemOpen
  \bibfield  {author} {\bibinfo {author} {\bibfnamefont {T.}~\bibnamefont
  {Hinderer}},\ }\bibfield  {title} {\bibinfo {title} {Tidal love numbers of
  neutron stars},\ }\href {https://doi.org/10.1086/533487} {\bibfield
  {journal} {\bibinfo  {journal} {The Astrophysical Journal}\ }\textbf
  {\bibinfo {volume} {677}},\ \bibinfo {pages} {1216} (\bibinfo {year}
  {2008})}\BibitemShut {NoStop}%
\bibitem [{\citenamefont {Damour}\ and\ \citenamefont
  {Nagar}(2009)}]{Damour:2009}%
  \BibitemOpen
  \bibfield  {author} {\bibinfo {author} {\bibfnamefont {T.}~\bibnamefont
  {Damour}}\ and\ \bibinfo {author} {\bibfnamefont {A.}~\bibnamefont {Nagar}},\
  }\bibfield  {title} {\bibinfo {title} {Relativistic tidal properties of
  neutron stars},\ }\href {https://doi.org/10.1103/PhysRevD.80.084035}
  {\bibfield  {journal} {\bibinfo  {journal} {Phys. Rev. D}\ }\textbf {\bibinfo
  {volume} {80}},\ \bibinfo {pages} {084035} (\bibinfo {year}
  {2009})}\BibitemShut {NoStop}%
\bibitem [{\citenamefont {Tolman}(1939)}]{Tolman:1939}%
  \BibitemOpen
  \bibfield  {author} {\bibinfo {author} {\bibfnamefont {R.~C.}\ \bibnamefont
  {Tolman}},\ }\href {https://doi.org/10.1103/PhysRev.55.364} {\bibfield
  {journal} {\bibinfo  {journal} {Phys. Rev.}\ }\textbf {\bibinfo {volume}
  {55}},\ \bibinfo {pages} {364} (\bibinfo {year} {1939})}\BibitemShut
  {NoStop}%
\bibitem [{\citenamefont {Oppenheimer}\ and\ \citenamefont
  {Volkoff}(1939)}]{Oppenheimer:1939}%
  \BibitemOpen
  \bibfield  {author} {\bibinfo {author} {\bibfnamefont {J.~R.}\ \bibnamefont
  {Oppenheimer}}\ and\ \bibinfo {author} {\bibfnamefont {G.~M.}\ \bibnamefont
  {Volkoff}},\ }\href {https://doi.org/10.1103/PhysRev.55.374} {\bibfield
  {journal} {\bibinfo  {journal} {Phys. Rev.}\ }\textbf {\bibinfo {volume}
  {55}},\ \bibinfo {pages} {374} (\bibinfo {year} {1939})}\BibitemShut
  {NoStop}%
\bibitem [{\citenamefont {Tews}\ \emph {et~al.}(2019)\citenamefont {Tews},
  \citenamefont {Margueron},\ and\ \citenamefont {Reddy}}]{Tews:2019}%
  \BibitemOpen
  \bibfield  {author} {\bibinfo {author} {\bibfnamefont {I.}~\bibnamefont
  {Tews}}, \bibinfo {author} {\bibfnamefont {J.}~\bibnamefont {Margueron}},\
  and\ \bibinfo {author} {\bibfnamefont {S.}~\bibnamefont {Reddy}},\ }\bibfield
   {title} {\bibinfo {title} {{Confronting gravitational-wave observations with
  modern nuclear physics constraints}},\ }\href
  {https://doi.org/10.1140/epja/i2019-12774-6} {\bibfield  {journal} {\bibinfo
  {journal} {European Journal of Physics A}\ }\textbf {\bibinfo {volume}
  {55}},\ \bibinfo {pages} {97} (\bibinfo {year} {2019})}\BibitemShut {NoStop}%
\bibitem [{\citenamefont {Xie}\ and\ \citenamefont {Li}(2021)}]{Xie:2021}%
  \BibitemOpen
  \bibfield  {author} {\bibinfo {author} {\bibfnamefont {W.-J.}\ \bibnamefont
  {Xie}}\ and\ \bibinfo {author} {\bibfnamefont {B.-A.}\ \bibnamefont {Li}},\
  }\bibfield  {title} {\bibinfo {title} {Bayesian inference of the dense-matter
  equation of state encapsulating a first-order hadron-quark phase transition
  from observables of canonical neutron stars},\ }\href
  {https://doi.org/10.1103/PhysRevC.103.035802} {\bibfield  {journal} {\bibinfo
   {journal} {Phys. Rev. C}\ }\textbf {\bibinfo {volume} {103}},\ \bibinfo
  {pages} {035802} (\bibinfo {year} {2021})}\BibitemShut {NoStop}%
\bibitem [{\citenamefont {Han}\ and\ \citenamefont {Steiner}(2019)}]{Han:2019}%
  \BibitemOpen
  \bibfield  {author} {\bibinfo {author} {\bibfnamefont {S.}~\bibnamefont
  {Han}}\ and\ \bibinfo {author} {\bibfnamefont {A.~W.}\ \bibnamefont
  {Steiner}},\ }\bibfield  {title} {\bibinfo {title} {Tidal deformability with
  sharp phase transitions in binary neutron stars},\ }\href
  {https://doi.org/10.1103/PhysRevD.99.083014} {\bibfield  {journal} {\bibinfo
  {journal} {Phys. Rev. D}\ }\textbf {\bibinfo {volume} {99}},\ \bibinfo
  {pages} {083014} (\bibinfo {year} {2019})}\BibitemShut {NoStop}%
\bibitem [{\citenamefont {Drischler}\ \emph {et~al.}(2019)\citenamefont
  {Drischler}, \citenamefont {Hebeler},\ and\ \citenamefont
  {Schwenk}}]{Drischler:2019}%
  \BibitemOpen
  \bibfield  {author} {\bibinfo {author} {\bibfnamefont {C.}~\bibnamefont
  {Drischler}}, \bibinfo {author} {\bibfnamefont {K.}~\bibnamefont {Hebeler}},\
  and\ \bibinfo {author} {\bibfnamefont {A.}~\bibnamefont {Schwenk}},\
  }\bibfield  {title} {\bibinfo {title} {Chiral interactions up to
  next-to-next-to-next-to-leading order and nuclear saturation},\ }\href
  {https://doi.org/10.1103/PhysRevLett.122.042501} {\bibfield  {journal}
  {\bibinfo  {journal} {Phys. Rev. Lett.}\ }\textbf {\bibinfo {volume} {122}},\
  \bibinfo {pages} {042501} (\bibinfo {year} {2019})}\BibitemShut {NoStop}%
\bibitem [{\citenamefont {Somasundaram}\ \emph {et~al.}(2021)\citenamefont
  {Somasundaram}, \citenamefont {Drischler}, \citenamefont {Tews},\ and\
  \citenamefont {Margueron}}]{Somasundaram:2021}%
  \BibitemOpen
  \bibfield  {author} {\bibinfo {author} {\bibfnamefont {R.}~\bibnamefont
  {Somasundaram}}, \bibinfo {author} {\bibfnamefont {C.}~\bibnamefont
  {Drischler}}, \bibinfo {author} {\bibfnamefont {I.}~\bibnamefont {Tews}},\
  and\ \bibinfo {author} {\bibfnamefont {J.}~\bibnamefont {Margueron}},\
  }\bibfield  {title} {\bibinfo {title} {Constraints on the nuclear symmetry
  energy from asymmetric-matter calculations with chiral $nn$ and $3n$
  interactions},\ }\href {https://doi.org/10.1103/PhysRevC.103.045803}
  {\bibfield  {journal} {\bibinfo  {journal} {Phys. Rev. C}\ }\textbf {\bibinfo
  {volume} {103}},\ \bibinfo {pages} {045803} (\bibinfo {year}
  {2021})}\BibitemShut {NoStop}%
\bibitem [{\citenamefont {Reinhard}\ \emph {et~al.}(2021)\citenamefont
  {Reinhard}, \citenamefont {Roca-Maza},\ and\ \citenamefont
  {Nazarewicz}}]{Reinhard:2021}%
  \BibitemOpen
  \bibfield  {author} {\bibinfo {author} {\bibfnamefont {P.-G.}\ \bibnamefont
  {Reinhard}}, \bibinfo {author} {\bibfnamefont {X.}~\bibnamefont
  {Roca-Maza}},\ and\ \bibinfo {author} {\bibfnamefont {W.}~\bibnamefont
  {Nazarewicz}},\ }\bibfield  {title} {\bibinfo {title} {{Information Content
  of the Parity-Violating Asymmetry in Pb208}},\ }\href
  {https://doi.org/10.1103/PhysRevLett.127.232501} {\bibfield  {journal}
  {\bibinfo  {journal} {Phys. Rev. Lett.}\ }\textbf {\bibinfo {volume} {127}},\
  \bibinfo {pages} {232501} (\bibinfo {year} {2021})},\ \Eprint
  {https://arxiv.org/abs/2105.15050} {arXiv:2105.15050 [nucl-th]} \BibitemShut
  {NoStop}%
\bibitem [{\citenamefont {Reed}\ \emph {et~al.}(2021)\citenamefont {Reed},
  \citenamefont {Fattoyev}, \citenamefont {Horowitz},\ and\ \citenamefont
  {Piekarewicz}}]{Reed:2021}%
  \BibitemOpen
  \bibfield  {author} {\bibinfo {author} {\bibfnamefont {B.~T.}\ \bibnamefont
  {Reed}}, \bibinfo {author} {\bibfnamefont {F.~J.}\ \bibnamefont {Fattoyev}},
  \bibinfo {author} {\bibfnamefont {C.~J.}\ \bibnamefont {Horowitz}},\ and\
  \bibinfo {author} {\bibfnamefont {J.}~\bibnamefont {Piekarewicz}},\
  }\bibfield  {title} {\bibinfo {title} {{Implications of PREX-2 on the
  Equation of State of Neutron-Rich Matter}},\ }\href
  {https://doi.org/10.1103/PhysRevLett.126.172503} {\bibfield  {journal}
  {\bibinfo  {journal} {Phys. Rev. Lett.}\ }\textbf {\bibinfo {volume} {126}},\
  \bibinfo {pages} {172503} (\bibinfo {year} {2021})},\ \Eprint
  {https://arxiv.org/abs/2101.03193} {arXiv:2101.03193 [nucl-th]} \BibitemShut
  {NoStop}%
\bibitem [{\citenamefont {Li}\ \emph {et~al.}(2023)\citenamefont {Li},
  \citenamefont {Sedrakian},\ and\ \citenamefont {Alford}}]{Li:2022}%
  \BibitemOpen
  \bibfield  {author} {\bibinfo {author} {\bibfnamefont {J.~J.}\ \bibnamefont
  {Li}}, \bibinfo {author} {\bibfnamefont {A.}~\bibnamefont {Sedrakian}},\ and\
  \bibinfo {author} {\bibfnamefont {M.}~\bibnamefont {Alford}},\ }\bibfield
  {title} {\bibinfo {title} {Ultracompact hybrid stars consistent with
  multimessenger astrophysics},\ }\href
  {https://doi.org/10.1103/PhysRevD.107.023018} {\bibfield  {journal} {\bibinfo
   {journal} {Phys. Rev. D}\ }\textbf {\bibinfo {volume} {107}},\ \bibinfo
  {pages} {023018} (\bibinfo {year} {2023})}\BibitemShut {NoStop}%
\bibitem [{\citenamefont {Capano}\ \emph {et~al.}(2020)\citenamefont {Capano},
  \citenamefont {Tews}, \citenamefont {Brown}, \citenamefont {Margalit},
  \citenamefont {De}, \citenamefont {Kumar}, \citenamefont {Brown},
  \citenamefont {Krishnan},\ and\ \citenamefont {Reddy}}]{Capano:2019eae}%
  \BibitemOpen
  \bibfield  {author} {\bibinfo {author} {\bibfnamefont {C.~D.}\ \bibnamefont
  {Capano}}, \bibinfo {author} {\bibfnamefont {I.}~\bibnamefont {Tews}},
  \bibinfo {author} {\bibfnamefont {S.~M.}\ \bibnamefont {Brown}}, \bibinfo
  {author} {\bibfnamefont {B.}~\bibnamefont {Margalit}}, \bibinfo {author}
  {\bibfnamefont {S.}~\bibnamefont {De}}, \bibinfo {author} {\bibfnamefont
  {S.}~\bibnamefont {Kumar}}, \bibinfo {author} {\bibfnamefont {D.~A.}\
  \bibnamefont {Brown}}, \bibinfo {author} {\bibfnamefont {B.}~\bibnamefont
  {Krishnan}},\ and\ \bibinfo {author} {\bibfnamefont {S.}~\bibnamefont
  {Reddy}},\ }\bibfield  {title} {\bibinfo {title} {{Stringent constraints on
  neutron-star radii from multimessenger observations and nuclear theory}},\
  }\href {https://doi.org/10.1038/s41550-020-1014-6} {\bibfield  {journal}
  {\bibinfo  {journal} {Nature Astron.}\ }\textbf {\bibinfo {volume} {4}},\
  \bibinfo {pages} {625} (\bibinfo {year} {2020})},\ \Eprint
  {https://arxiv.org/abs/1908.10352} {arXiv:1908.10352 [astro-ph.HE]}
  \BibitemShut {NoStop}%
\bibitem [{\citenamefont {G\"uven}\ \emph {et~al.}(2020)\citenamefont
  {G\"uven}, \citenamefont {Bozkurt}, \citenamefont {Khan},\ and\ \citenamefont
  {Margueron}}]{Guven:2020}%
  \BibitemOpen
  \bibfield  {author} {\bibinfo {author} {\bibfnamefont {H.}~\bibnamefont
  {G\"uven}}, \bibinfo {author} {\bibfnamefont {K.}~\bibnamefont {Bozkurt}},
  \bibinfo {author} {\bibfnamefont {E.}~\bibnamefont {Khan}},\ and\ \bibinfo
  {author} {\bibfnamefont {J.}~\bibnamefont {Margueron}},\ }\bibfield  {title}
  {\bibinfo {title} {{Multi-messenger and multi-physics Bayesian inference for
  GW170817 binary neutron star merger}},\ }\href
  {https://doi.org/10.1103/PhysRevC.102.015805} {\bibfield  {journal} {\bibinfo
   {journal} {Phys. Rev. C}\ }\textbf {\bibinfo {volume} {102}},\ \bibinfo
  {pages} {015805} (\bibinfo {year} {2020})},\ \Eprint
  {https://arxiv.org/abs/2001.10259} {arXiv:2001.10259 [nucl-th]} \BibitemShut
  {NoStop}%
\bibitem [{\citenamefont {Pang}\ \emph {et~al.}(2021)\citenamefont {Pang},
  \citenamefont {Tews}, \citenamefont {Coughlin}, \citenamefont {Bulla},
  \citenamefont {Broeck},\ and\ \citenamefont {Dietrich}}]{Pang:2021}%
  \BibitemOpen
  \bibfield  {author} {\bibinfo {author} {\bibfnamefont {P.~T.~H.}\
  \bibnamefont {Pang}}, \bibinfo {author} {\bibfnamefont {I.}~\bibnamefont
  {Tews}}, \bibinfo {author} {\bibfnamefont {M.~W.}\ \bibnamefont {Coughlin}},
  \bibinfo {author} {\bibfnamefont {M.}~\bibnamefont {Bulla}}, \bibinfo
  {author} {\bibfnamefont {C.~V.~D.}\ \bibnamefont {Broeck}},\ and\ \bibinfo
  {author} {\bibfnamefont {T.}~\bibnamefont {Dietrich}},\ }\href
  {https://doi.org/10.3847/1538-4357/ac19ab} {\bibfield  {journal} {\bibinfo
  {journal} {The Astrophysical Journal}\ }\textbf {\bibinfo {volume} {922}},\
  \bibinfo {pages} {14} (\bibinfo {year} {2021})}\BibitemShut {NoStop}%
\bibitem [{\citenamefont {Dinh~Thi}\ \emph {et~al.}(2021)\citenamefont
  {Dinh~Thi}, \citenamefont {Mondal},\ and\ \citenamefont
  {Gulminelli}}]{Dinh:2021}%
  \BibitemOpen
  \bibfield  {author} {\bibinfo {author} {\bibfnamefont {H.}~\bibnamefont
  {Dinh~Thi}}, \bibinfo {author} {\bibfnamefont {C.}~\bibnamefont {Mondal}},\
  and\ \bibinfo {author} {\bibfnamefont {F.}~\bibnamefont {Gulminelli}},\
  }\bibfield  {title} {\bibinfo {title} {The nuclear matter density functional
  under the nucleonic hypothesis},\ }\bibfield  {journal} {\bibinfo  {journal}
  {Universe}\ }\textbf {\bibinfo {volume} {7}},\ \href
  {https://doi.org/10.3390/universe7100373} {10.3390/universe7100373} (\bibinfo
  {year} {2021})\BibitemShut {NoStop}%
\bibitem [{\citenamefont {Pankow}(2018)}]{Pankow:2018}%
  \BibitemOpen
  \bibfield  {author} {\bibinfo {author} {\bibfnamefont {C.}~\bibnamefont
  {Pankow}},\ }\bibfield  {title} {\bibinfo {title} {On gw170817 and the
  galactic binary neutron star population},\ }\href
  {https://doi.org/10.3847/1538-4357/aadc66} {\bibfield  {journal} {\bibinfo
  {journal} {The Astrophysical Journal}\ }\textbf {\bibinfo {volume} {866}},\
  \bibinfo {pages} {60} (\bibinfo {year} {2018})}\BibitemShut {NoStop}%
\bibitem [{\citenamefont {Horvath}(2019)}]{Horvath:2019}%
  \BibitemOpen
  \bibfield  {author} {\bibinfo {author} {\bibfnamefont {J.~E.}\ \bibnamefont
  {Horvath}},\ }\bibfield  {title} {\bibinfo {title} {{The binaries of the
  NS-NS merging events}},\ }\href {https://doi.org/10.1063/1.5117805}
  {\bibfield  {journal} {\bibinfo  {journal} {AIP Conference Proceedings}\
  }\textbf {\bibinfo {volume} {2127}},\ \bibinfo {pages} {020015} (\bibinfo
  {year} {2019})},\ \Eprint
  {https://arxiv.org/abs/https://pubs.aip.org/aip/acp/article-pdf/doi/10.1063/1.5117805/14067429/020015\_1\_online.pdf}
  {https://pubs.aip.org/aip/acp/article-pdf/doi/10.1063/1.5117805/14067429/020015\_1\_online.pdf}
  \BibitemShut {NoStop}%
\bibitem [{\citenamefont {Ferdman}\ \emph {et~al.}(2020)\citenamefont
  {Ferdman}, \citenamefont {Freire}, \citenamefont {Perera}, \citenamefont
  {Pol}, \citenamefont {Camilo}, \citenamefont {Chatterjee}, \citenamefont
  {Cordes}, \citenamefont {Crawford}, \citenamefont {Hessels}, \citenamefont
  {Kaspi} \emph {et~al.}}]{Ferdman:2020}%
  \BibitemOpen
  \bibfield  {author} {\bibinfo {author} {\bibfnamefont {R.}~\bibnamefont
  {Ferdman}}, \bibinfo {author} {\bibfnamefont {P.}~\bibnamefont {Freire}},
  \bibinfo {author} {\bibfnamefont {B.}~\bibnamefont {Perera}}, \bibinfo
  {author} {\bibfnamefont {N.}~\bibnamefont {Pol}}, \bibinfo {author}
  {\bibfnamefont {F.}~\bibnamefont {Camilo}}, \bibinfo {author} {\bibfnamefont
  {S.}~\bibnamefont {Chatterjee}}, \bibinfo {author} {\bibfnamefont
  {J.}~\bibnamefont {Cordes}}, \bibinfo {author} {\bibfnamefont
  {F.}~\bibnamefont {Crawford}}, \bibinfo {author} {\bibfnamefont
  {J.}~\bibnamefont {Hessels}}, \bibinfo {author} {\bibfnamefont
  {V.}~\bibnamefont {Kaspi}}, \emph {et~al.},\ }\bibfield  {title} {\bibinfo
  {title} {Asymmetric mass ratios for bright double neutron-star mergers},\
  }\href@noop {} {\bibfield  {journal} {\bibinfo  {journal} {Nature}\ }\textbf
  {\bibinfo {volume} {583}},\ \bibinfo {pages} {211} (\bibinfo {year}
  {2020})}\BibitemShut {NoStop}%
\bibitem [{\citenamefont {Chodos}\ \emph {et~al.}(1974)\citenamefont {Chodos},
  \citenamefont {Jaffe}, \citenamefont {Johnson}, \citenamefont {Thorn},\ and\
  \citenamefont {Weisskopf}}]{Chodos:1974}%
  \BibitemOpen
  \bibfield  {author} {\bibinfo {author} {\bibfnamefont {A.}~\bibnamefont
  {Chodos}}, \bibinfo {author} {\bibfnamefont {R.~L.}\ \bibnamefont {Jaffe}},
  \bibinfo {author} {\bibfnamefont {K.}~\bibnamefont {Johnson}}, \bibinfo
  {author} {\bibfnamefont {C.~B.}\ \bibnamefont {Thorn}},\ and\ \bibinfo
  {author} {\bibfnamefont {V.~F.}\ \bibnamefont {Weisskopf}},\ }\bibfield
  {title} {\bibinfo {title} {New extended model of hadrons},\ }\href
  {https://doi.org/10.1103/PhysRevD.9.3471} {\bibfield  {journal} {\bibinfo
  {journal} {Phys. Rev. D}\ }\textbf {\bibinfo {volume} {9}},\ \bibinfo {pages}
  {3471} (\bibinfo {year} {1974})}\BibitemShut {NoStop}%
\bibitem [{\citenamefont {{Zdunik, J. L.}}\ and\ \citenamefont {{Haensel,
  P.}}(2013)}]{Zdunik:2013}%
  \BibitemOpen
  \bibfield  {author} {\bibinfo {author} {\bibnamefont {{Zdunik, J. L.}}}\ and\
  \bibinfo {author} {\bibnamefont {{Haensel, P.}}},\ }\bibfield  {title}
  {\bibinfo {title} {Maximum mass of neutron stars and strange neutron-star
  cores},\ }\href {https://doi.org/10.1051/0004-6361/201220697} {\bibfield
  {journal} {\bibinfo  {journal} {A\&A}\ }\textbf {\bibinfo {volume} {551}},\
  \bibinfo {pages} {A61} (\bibinfo {year} {2013})}\BibitemShut {NoStop}%
\bibitem [{\citenamefont {Alford}\ \emph {et~al.}(2013)\citenamefont {Alford},
  \citenamefont {Han},\ and\ \citenamefont {Prakash}}]{Alford:2013}%
  \BibitemOpen
  \bibfield  {author} {\bibinfo {author} {\bibfnamefont {M.~G.}\ \bibnamefont
  {Alford}}, \bibinfo {author} {\bibfnamefont {S.}~\bibnamefont {Han}},\ and\
  \bibinfo {author} {\bibfnamefont {M.}~\bibnamefont {Prakash}},\ }\bibfield
  {title} {\bibinfo {title} {{Generic conditions for stable hybrid stars}},\
  }\href {https://doi.org/10.1103/PhysRevD.88.083013} {\bibfield  {journal}
  {\bibinfo  {journal} {Phys. Rev. D}\ }\textbf {\bibinfo {volume} {88}},\
  \bibinfo {pages} {083013} (\bibinfo {year} {2013})},\ \Eprint
  {https://arxiv.org/abs/1302.4732} {arXiv:1302.4732 [astro-ph.SR]}
  \BibitemShut {NoStop}%
\bibitem [{\citenamefont {{Chamel, N.}}\ \emph {et~al.}(2013)\citenamefont
  {{Chamel, N.}}, \citenamefont {{Fantina, A. F.}}, \citenamefont {{Pearson, J.
  M.}},\ and\ \citenamefont {{Goriely, S.}}}]{Chamel:2013}%
  \BibitemOpen
  \bibfield  {author} {\bibinfo {author} {\bibnamefont {{Chamel, N.}}},
  \bibinfo {author} {\bibnamefont {{Fantina, A. F.}}}, \bibinfo {author}
  {\bibnamefont {{Pearson, J. M.}}},\ and\ \bibinfo {author} {\bibnamefont
  {{Goriely, S.}}},\ }\bibfield  {title} {\bibinfo {title} {Phase transitions
  in dense matter and the maximum mass of neutron stars},\ }\href
  {https://doi.org/10.1051/0004-6361/201220986} {\bibfield  {journal} {\bibinfo
   {journal} {A\&A}\ }\textbf {\bibinfo {volume} {553}},\ \bibinfo {pages}
  {A22} (\bibinfo {year} {2013})}\BibitemShut {NoStop}%
\bibitem [{\citenamefont {Somasundaram}\ and\ \citenamefont
  {Margueron}(2022)}]{Somasundaram:2022a}%
  \BibitemOpen
  \bibfield  {author} {\bibinfo {author} {\bibfnamefont {R.}~\bibnamefont
  {Somasundaram}}\ and\ \bibinfo {author} {\bibfnamefont {J.}~\bibnamefont
  {Margueron}},\ }\bibfield  {title} {\bibinfo {title} {{Impact of massive
  neutron star radii on the nature of phase transitions in dense matter}},\
  }\href {https://doi.org/10.1209/0295-5075/ac63de} {\bibfield  {journal}
  {\bibinfo  {journal} {EPL}\ }\textbf {\bibinfo {volume} {138}},\ \bibinfo
  {pages} {14002} (\bibinfo {year} {2022})},\ \Eprint
  {https://arxiv.org/abs/2104.13612} {arXiv:2104.13612 [astro-ph.HE]}
  \BibitemShut {NoStop}%
\bibitem [{\citenamefont {Somasundaram}\ \emph {et~al.}(2023)\citenamefont
  {Somasundaram}, \citenamefont {Tews},\ and\ \citenamefont
  {Margueron}}]{Somasundaram:2022b}%
  \BibitemOpen
  \bibfield  {author} {\bibinfo {author} {\bibfnamefont {R.}~\bibnamefont
  {Somasundaram}}, \bibinfo {author} {\bibfnamefont {I.}~\bibnamefont {Tews}},\
  and\ \bibinfo {author} {\bibfnamefont {J.}~\bibnamefont {Margueron}},\
  }\bibfield  {title} {\bibinfo {title} {Investigating signatures of phase
  transitions in neutron-star cores},\ }\href
  {https://doi.org/10.1103/PhysRevC.107.025801} {\bibfield  {journal} {\bibinfo
   {journal} {Phys. Rev. C}\ }\textbf {\bibinfo {volume} {107}},\ \bibinfo
  {pages} {025801} (\bibinfo {year} {2023})},\ \Eprint
  {https://arxiv.org/abs/2112.08157} {arXiv:2112.08157 [nucl-th]} \BibitemShut
  {NoStop}%
\bibitem [{\citenamefont {Alford}\ and\ \citenamefont
  {Sedrakian}(2017)}]{Alford:2017}%
  \BibitemOpen
  \bibfield  {author} {\bibinfo {author} {\bibfnamefont {M.}~\bibnamefont
  {Alford}}\ and\ \bibinfo {author} {\bibfnamefont {A.}~\bibnamefont
  {Sedrakian}},\ }\bibfield  {title} {\bibinfo {title} {Compact stars with
  sequential qcd phase transitions},\ }\href
  {https://doi.org/10.1103/PhysRevLett.119.161104} {\bibfield  {journal}
  {\bibinfo  {journal} {Phys. Rev. Lett.}\ }\textbf {\bibinfo {volume} {119}},\
  \bibinfo {pages} {161104} (\bibinfo {year} {2017})}\BibitemShut {NoStop}%
\bibitem [{\citenamefont {Agrawal}(2010)}]{Agrawal:2010}%
  \BibitemOpen
  \bibfield  {author} {\bibinfo {author} {\bibfnamefont {B.~K.}\ \bibnamefont
  {Agrawal}},\ }\bibfield  {title} {\bibinfo {title} {Equations of state and
  stability of color-superconducting quark matter cores in hybrid stars},\
  }\href {https://doi.org/10.1103/PhysRevD.81.023009} {\bibfield  {journal}
  {\bibinfo  {journal} {Phys. Rev. D}\ }\textbf {\bibinfo {volume} {81}},\
  \bibinfo {pages} {023009} (\bibinfo {year} {2010})}\BibitemShut {NoStop}%
\bibitem [{\citenamefont {Pfaff}\ \emph {et~al.}(2022)\citenamefont {Pfaff},
  \citenamefont {Hansen},\ and\ \citenamefont {Gulminelli}}]{Pfaff:2022}%
  \BibitemOpen
  \bibfield  {author} {\bibinfo {author} {\bibfnamefont {A.}~\bibnamefont
  {Pfaff}}, \bibinfo {author} {\bibfnamefont {H.}~\bibnamefont {Hansen}},\ and\
  \bibinfo {author} {\bibfnamefont {F.}~\bibnamefont {Gulminelli}},\ }\bibfield
   {title} {\bibinfo {title} {Bayesian analysis of the properties of hybrid
  stars with the nambu--jona-lasinio model},\ }\href
  {https://doi.org/10.1103/PhysRevC.105.035802} {\bibfield  {journal} {\bibinfo
   {journal} {Phys. Rev. C}\ }\textbf {\bibinfo {volume} {105}},\ \bibinfo
  {pages} {035802} (\bibinfo {year} {2022})}\BibitemShut {NoStop}%
\bibitem [{\citenamefont {Ayriyan}\ \emph {et~al.}(2019)\citenamefont
  {Ayriyan}, \citenamefont {Alvarez-Castillo}, \citenamefont {Blaschke},\ and\
  \citenamefont {Grigorian}}]{Ayriyan:2019}%
  \BibitemOpen
  \bibfield  {author} {\bibinfo {author} {\bibfnamefont {A.}~\bibnamefont
  {Ayriyan}}, \bibinfo {author} {\bibfnamefont {D.}~\bibnamefont
  {Alvarez-Castillo}}, \bibinfo {author} {\bibfnamefont {D.}~\bibnamefont
  {Blaschke}},\ and\ \bibinfo {author} {\bibfnamefont {H.}~\bibnamefont
  {Grigorian}},\ }\bibfield  {title} {\bibinfo {title} {Bayesian analysis for
  extracting properties of the nuclear equation of state from observational
  data including tidal deformability from gw170817},\ }\bibfield  {journal}
  {\bibinfo  {journal} {Universe}\ }\textbf {\bibinfo {volume} {5}},\ \href
  {https://doi.org/10.3390/universe5020061} {10.3390/universe5020061} (\bibinfo
  {year} {2019})\BibitemShut {NoStop}%
\bibitem [{\citenamefont {Parisi}\ \emph {et~al.}(2021)\citenamefont {Parisi},
  \citenamefont {Flores}, \citenamefont {Lenzi}, \citenamefont {Chen},\ and\
  \citenamefont {Lugones}}]{Parisi:2021}%
  \BibitemOpen
  \bibfield  {author} {\bibinfo {author} {\bibfnamefont {A.}~\bibnamefont
  {Parisi}}, \bibinfo {author} {\bibfnamefont {C.~V.}\ \bibnamefont {Flores}},
  \bibinfo {author} {\bibfnamefont {C.~H.}\ \bibnamefont {Lenzi}}, \bibinfo
  {author} {\bibfnamefont {C.-S.}\ \bibnamefont {Chen}},\ and\ \bibinfo
  {author} {\bibfnamefont {G.}~\bibnamefont {Lugones}},\ }\bibfield  {title}
  {\bibinfo {title} {Hybrid stars in the light of the merging event
  {GW}170817},\ }\href {https://doi.org/10.1088/1475-7516/2021/06/042}
  {\bibfield  {journal} {\bibinfo  {journal} {Journal of Cosmology and
  Astroparticle Physics}\ }\textbf {\bibinfo {volume} {2021}}\bibinfo  {number}
  { (06)},\ \bibinfo {pages} {042}}\BibitemShut {NoStop}%
\bibitem [{\citenamefont {Steiner}(2015)}]{Steiner:2015}%
  \BibitemOpen
\bibfield  {number} {  }\bibfield  {author} {\bibinfo {author} {\bibfnamefont
  {A.~W.}\ \bibnamefont {Steiner}},\ }\bibfield  {title} {\bibinfo {title}
  {Moving beyond chi-squared in nuclei and neutron stars},\ }\href
  {https://doi.org/10.1088/0954-3899/42/3/034004} {\bibfield  {journal}
  {\bibinfo  {journal} {Journal of Physics G: Nuclear and Particle Physics}\
  }\textbf {\bibinfo {volume} {42}},\ \bibinfo {pages} {034004} (\bibinfo
  {year} {2015})}\BibitemShut {NoStop}%
\bibitem [{\citenamefont {Dietrich}\ \emph {et~al.}(2020)\citenamefont
  {Dietrich}, \citenamefont {Coughlin}, \citenamefont {Pang}, \citenamefont
  {Bulla}, \citenamefont {Heinzel}, \citenamefont {Issa}, \citenamefont
  {Tews},\ and\ \citenamefont {Antier}}]{Dietrich:2020}%
  \BibitemOpen
  \bibfield  {author} {\bibinfo {author} {\bibfnamefont {T.}~\bibnamefont
  {Dietrich}}, \bibinfo {author} {\bibfnamefont {M.~W.}\ \bibnamefont
  {Coughlin}}, \bibinfo {author} {\bibfnamefont {P.~T.~H.}\ \bibnamefont
  {Pang}}, \bibinfo {author} {\bibfnamefont {M.}~\bibnamefont {Bulla}},
  \bibinfo {author} {\bibfnamefont {J.}~\bibnamefont {Heinzel}}, \bibinfo
  {author} {\bibfnamefont {L.}~\bibnamefont {Issa}}, \bibinfo {author}
  {\bibfnamefont {I.}~\bibnamefont {Tews}},\ and\ \bibinfo {author}
  {\bibfnamefont {S.}~\bibnamefont {Antier}},\ }\bibfield  {title} {\bibinfo
  {title} {Multimessenger constraints on the neutron-star equation of state and
  the hubble constant},\ }\href {https://doi.org/10.1126/science.abb4317}
  {\bibfield  {journal} {\bibinfo  {journal} {Science}\ }\textbf {\bibinfo
  {volume} {370}},\ \bibinfo {pages} {1450} (\bibinfo {year} {2020})},\ \Eprint
  {https://arxiv.org/abs/https://www.science.org/doi/pdf/10.1126/science.abb4317}
  {https://www.science.org/doi/pdf/10.1126/science.abb4317} \BibitemShut
  {NoStop}%
\bibitem [{\citenamefont {Al-Mamun}\ \emph {et~al.}(2021)\citenamefont
  {Al-Mamun}, \citenamefont {Steiner}, \citenamefont {N\"attil\"a},
  \citenamefont {Lange}, \citenamefont {O'Shaughnessy}, \citenamefont {Tews},
  \citenamefont {Gandolfi}, \citenamefont {Heinke},\ and\ \citenamefont
  {Han}}]{Al-Mamun:2021}%
  \BibitemOpen
  \bibfield  {author} {\bibinfo {author} {\bibfnamefont {M.}~\bibnamefont
  {Al-Mamun}}, \bibinfo {author} {\bibfnamefont {A.~W.}\ \bibnamefont
  {Steiner}}, \bibinfo {author} {\bibfnamefont {J.}~\bibnamefont
  {N\"attil\"a}}, \bibinfo {author} {\bibfnamefont {J.}~\bibnamefont {Lange}},
  \bibinfo {author} {\bibfnamefont {R.}~\bibnamefont {O'Shaughnessy}}, \bibinfo
  {author} {\bibfnamefont {I.}~\bibnamefont {Tews}}, \bibinfo {author}
  {\bibfnamefont {S.}~\bibnamefont {Gandolfi}}, \bibinfo {author}
  {\bibfnamefont {C.}~\bibnamefont {Heinke}},\ and\ \bibinfo {author}
  {\bibfnamefont {S.}~\bibnamefont {Han}},\ }\bibfield  {title} {\bibinfo
  {title} {Combining electromagnetic and gravitational-wave constraints on
  neutron-star masses and radii},\ }\href
  {https://doi.org/10.1103/PhysRevLett.126.061101} {\bibfield  {journal}
  {\bibinfo  {journal} {Phys. Rev. Lett.}\ }\textbf {\bibinfo {volume} {126}},\
  \bibinfo {pages} {061101} (\bibinfo {year} {2021})}\BibitemShut {NoStop}%
\end{thebibliography}%


\begin{thebibliography}{4}%
\makeatletter
\providecommand \@ifxundefined [1]{%
 \@ifx{#1\undefined}
}%
\providecommand \@ifnum [1]{%
 \ifnum #1\expandafter \@firstoftwo
 \else \expandafter \@secondoftwo
 \fi
}%
\providecommand \@ifx [1]{%
 \ifx #1\expandafter \@firstoftwo
 \else \expandafter \@secondoftwo
 \fi
}%
\providecommand \natexlab [1]{#1}%
\providecommand \enquote  [1]{``#1''}%
\providecommand \bibnamefont  [1]{#1}%
\providecommand \bibfnamefont [1]{#1}%
\providecommand \citenamefont [1]{#1}%
\providecommand \href@noop [0]{\@secondoftwo}%
\providecommand \href [0]{\begingroup \@sanitize@url \@href}%
\providecommand \@href[1]{\@@startlink{#1}\@@href}%
\providecommand \@@href[1]{\endgroup#1\@@endlink}%
\providecommand \@sanitize@url [0]{\catcode `\\12\catcode `\$12\catcode
  `\&12\catcode `\#12\catcode `\^12\catcode `\_12\catcode `\%12\relax}%
\providecommand \@@startlink[1]{}%
\providecommand \@@endlink[0]{}%
\providecommand \url  [0]{\begingroup\@sanitize@url \@url }%
\providecommand \@url [1]{\endgroup\@href {#1}{\urlprefix }}%
\providecommand \urlprefix  [0]{URL }%
\providecommand \Eprint [0]{\href }%
\providecommand \doibase [0]{http://dx.doi.org/}%
\providecommand \selectlanguage [0]{\@gobble}%
\providecommand \bibinfo  [0]{\@secondoftwo}%
\providecommand \bibfield  [0]{\@secondoftwo}%
\providecommand \translation [1]{[#1]}%
\providecommand \BibitemOpen [0]{}%
\providecommand \bibitemStop [0]{}%
\providecommand \bibitemNoStop [0]{.\EOS\space}%
\providecommand \EOS [0]{\spacefactor3000\relax}%
\providecommand \BibitemShut  [1]{\csname bibitem#1\endcsname}%
\let\auto@bib@innerbib\@empty
\bibitem [{\citenamefont {Abbott}\ \emph {et~al.}(2019)\citenamefont {Abbott},
  \citenamefont {Abbott}, \citenamefont {Abbott} \emph {et~al.}}]{Abbott:2019}%
  \BibitemOpen
  \bibfield  {author} {\bibinfo {author} {\bibfnamefont {B.~P.}\ \bibnamefont
  {Abbott}}, \bibinfo {author} {\bibfnamefont {R.}~\bibnamefont {Abbott}},
  \bibinfo {author} {\bibfnamefont {T.~D.}\ \bibnamefont {Abbott}},  \emph
  {et~al.} (\bibinfo {collaboration} {LIGO Scientific Collaboration and Virgo
  Collaboration}),\ }\href {\doibase 10.1103/PhysRevX.9.011001} {\bibfield
  {journal} {\bibinfo  {journal} {Phys. Rev. X}\ }\textbf {\bibinfo {volume}
  {9}},\ \bibinfo {pages} {011001} (\bibinfo {year} {2019})}\BibitemShut
  {NoStop}%
\bibitem [{\citenamefont {Flanagan}\ and\ \citenamefont
  {Hinderer}(2008)}]{Flanagan:2008}%
  \BibitemOpen
  \bibfield  {author} {\bibinfo {author} {\bibfnamefont {E.~E.}\ \bibnamefont
  {Flanagan}}\ and\ \bibinfo {author} {\bibfnamefont {T.}~\bibnamefont
  {Hinderer}},\ }\href {\doibase 10.1103/PhysRevD.77.021502} {\bibfield
  {journal} {\bibinfo  {journal} {Phys. Rev. D}\ }\textbf {\bibinfo {volume}
  {77}},\ \bibinfo {pages} {021502(R)} (\bibinfo {year} {2008})}\BibitemShut
  {NoStop}%
\bibitem [{\citenamefont {Damour}\ and\ \citenamefont
  {Nagar}(2009)}]{Damour:2009}%
  \BibitemOpen
  \bibfield  {author} {\bibinfo {author} {\bibfnamefont {T.}~\bibnamefont
  {Damour}}\ and\ \bibinfo {author} {\bibfnamefont {A.}~\bibnamefont {Nagar}},\
  }\href {\doibase 10.1103/PhysRevD.80.084035} {\bibfield  {journal} {\bibinfo
  {journal} {Phys. Rev. D}\ }\textbf {\bibinfo {volume} {80}},\ \bibinfo
  {pages} {084035} (\bibinfo {year} {2009})}\BibitemShut {NoStop}%
\bibitem [{\citenamefont {De}\ \emph {et~al.}(2018)\citenamefont {De},
  \citenamefont {Finstad}, \citenamefont {Lattimer}, \citenamefont {Brown},
  \citenamefont {Berger},\ and\ \citenamefont {Biwer}}]{De:2018}%
  \BibitemOpen
  \bibfield  {author} {\bibinfo {author} {\bibfnamefont {S.}~\bibnamefont
  {De}}, \bibinfo {author} {\bibfnamefont {D.}~\bibnamefont {Finstad}},
  \bibinfo {author} {\bibfnamefont {J.~M.}\ \bibnamefont {Lattimer}}, \bibinfo
  {author} {\bibfnamefont {D.~A.}\ \bibnamefont {Brown}}, \bibinfo {author}
  {\bibfnamefont {E.}~\bibnamefont {Berger}}, \ and\ \bibinfo {author}
  {\bibfnamefont {C.~M.}\ \bibnamefont {Biwer}},\ }\href {\doibase
  10.1103/PhysRevLett.121.091102} {\bibfield  {journal} {\bibinfo  {journal}
  {Phys. Rev. Lett.}\ }\textbf {\bibinfo {volume} {121}},\ \bibinfo {pages}
  {091102} (\bibinfo {year} {2018})}\BibitemShut {NoStop}%
\end{thebibliography}%

\end{document}


%

\title{Supplemental material: On the nature of compact stars determined by gravitational waves, radio-astronomy, x-ray emission and nuclear physics}


\author{H. G\"{u}ven}
\affiliation{Universit\'e Paris-Saclay, CNRS/IN2P3, IJCLab, 91405 Orsay, France}
\affiliation{Physics Department, Yildiz Technical University, 34220 Esenler, Istanbul, Turkey}

\author{J. Margueron}
\affiliation{Univ Lyon, Univ Claude Bernard Lyon 1, CNRS/IN2P3, IP2I Lyon, UMR 5822, F-69622, Villeurbanne, France}
\date{\today}

\author{K. Bozkurt}
\affiliation{Physics Department, Yildiz Technical University, 34220 Esenler, Istanbul, Turkey}
\affiliation{Universit\'e Paris-Saclay, CNRS/IN2P3, IJCLab, 91405 Orsay, France}

\author{E. Khan}
\affiliation{Universit\'e Paris-Saclay, CNRS/IN2P3, IJCLab, 91405 Orsay, France}
\affiliation{Institut Universitaire de France (IUF)}

\begin{abstract}
In this supplemental material, we provide additional figures exploring the impact of the masses in the binary system and of the distribution of the tidal deformability in our analysis.
\end{abstract}

\maketitle

\begin{figure}[t]
\centering
\includegraphics[width=0.48\textwidth]{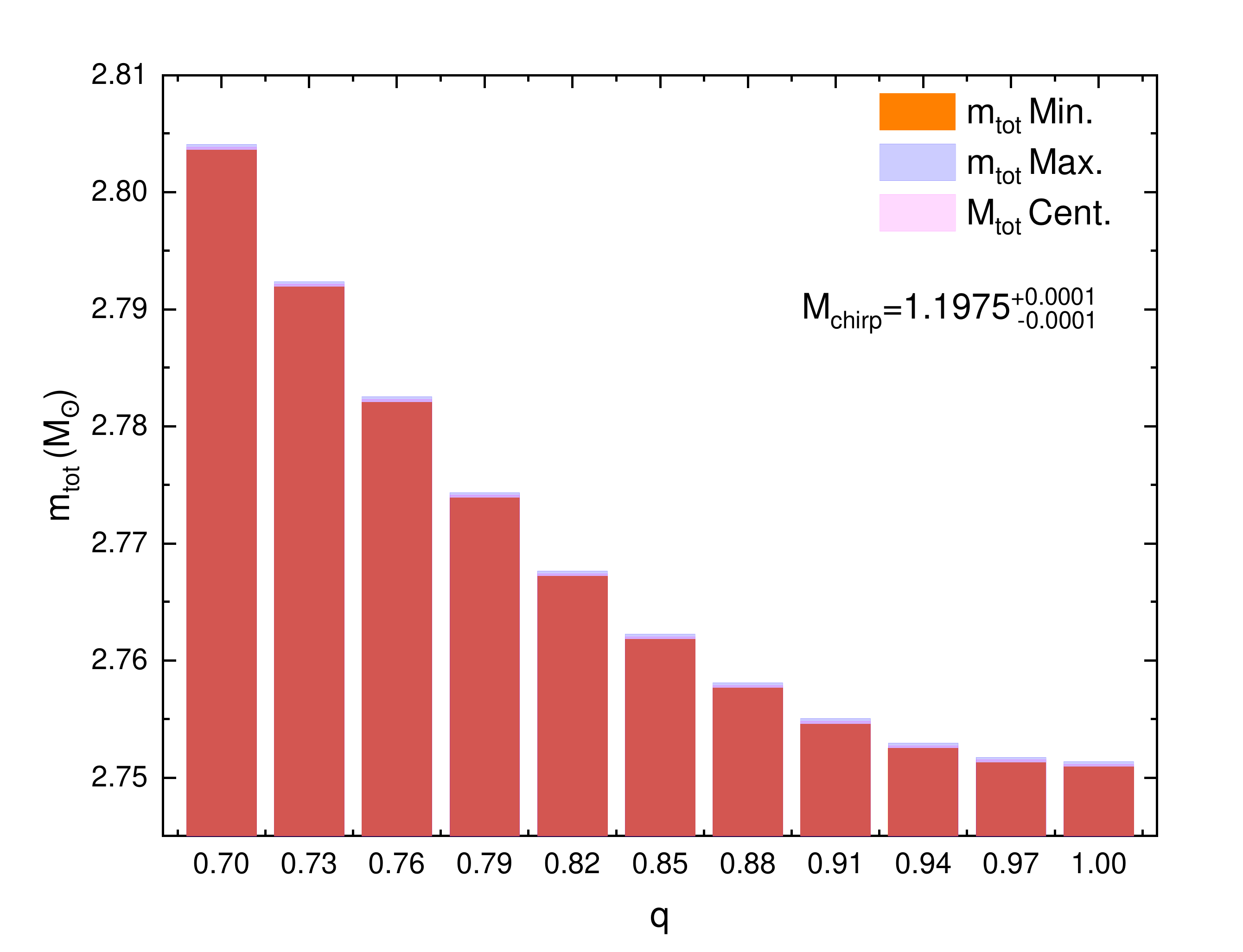}
\caption{The total mass $m_\mathrm{tot}$ of the binary system as function of the mass ratio $q$, where the chirp mass is fixed to be $M_\mathrm{chirp}=1.1975(1)~\mathrm{M}_\odot$~\cite{Abbott:2019}. The dependence of $m_\mathrm{tot}$ in the mass ratio $q$ shall be compared to the uncertainty in $m_\mathrm{tot}$ which is considered in the second prescription, see text for more details.}
\label{fig:Mchirp}
\end{figure}

The distributions of masses $m_1$ and $m_2$ defining the tidal deformability have been obtained in our analysis from the chirp mass $M_\mathrm{chirp}=(m_1 m_2)^{3/5}m_\mathrm{tot}^{-1/5}$~\cite{Abbott:2019}, $m_\mathrm{tot}=m_1+m_2$, and the mass fraction $q=m_1/m_2$ in the interval $[0.73-1]$ at 90\% confidence level. It can also be extracted assuming a constant value of the total mass $m_\mathrm{tot}=2.73^{+0.03}_{-0.04}~\mathrm{M}_\odot$ independently of the mass ratio $q$. These two prescriptions are not totally equivalent as one can see from Fig.~\ref{fig:Mchirp}. The total mass $m_\mathrm{tot}$ is represented as function of $q$, where $m_1$ and $m_2$ are obtained from $M_\mathrm{chirp}$ and $q=[0.7,1]$. Fig.~\ref{fig:Mchirp} shows that a constant value of $M_\mathrm{chirp}$ implies a non-constant distribution of $m_\mathrm{tot}$, which decreases by 0.05M$_\odot$ as $q$ varies from 0.7 up to 1.
One shall indeed prefer this prescription since $M_\mathrm{chirp}$ is directly extracted from the data: $M_\mathrm{chirp}$ is the third-order post-Newtonian correction to the wavefront phase~\cite{Flanagan:2008,Damour:2009}.

\begin{figure}[t]
\centering
\includegraphics[width=0.48\textwidth]{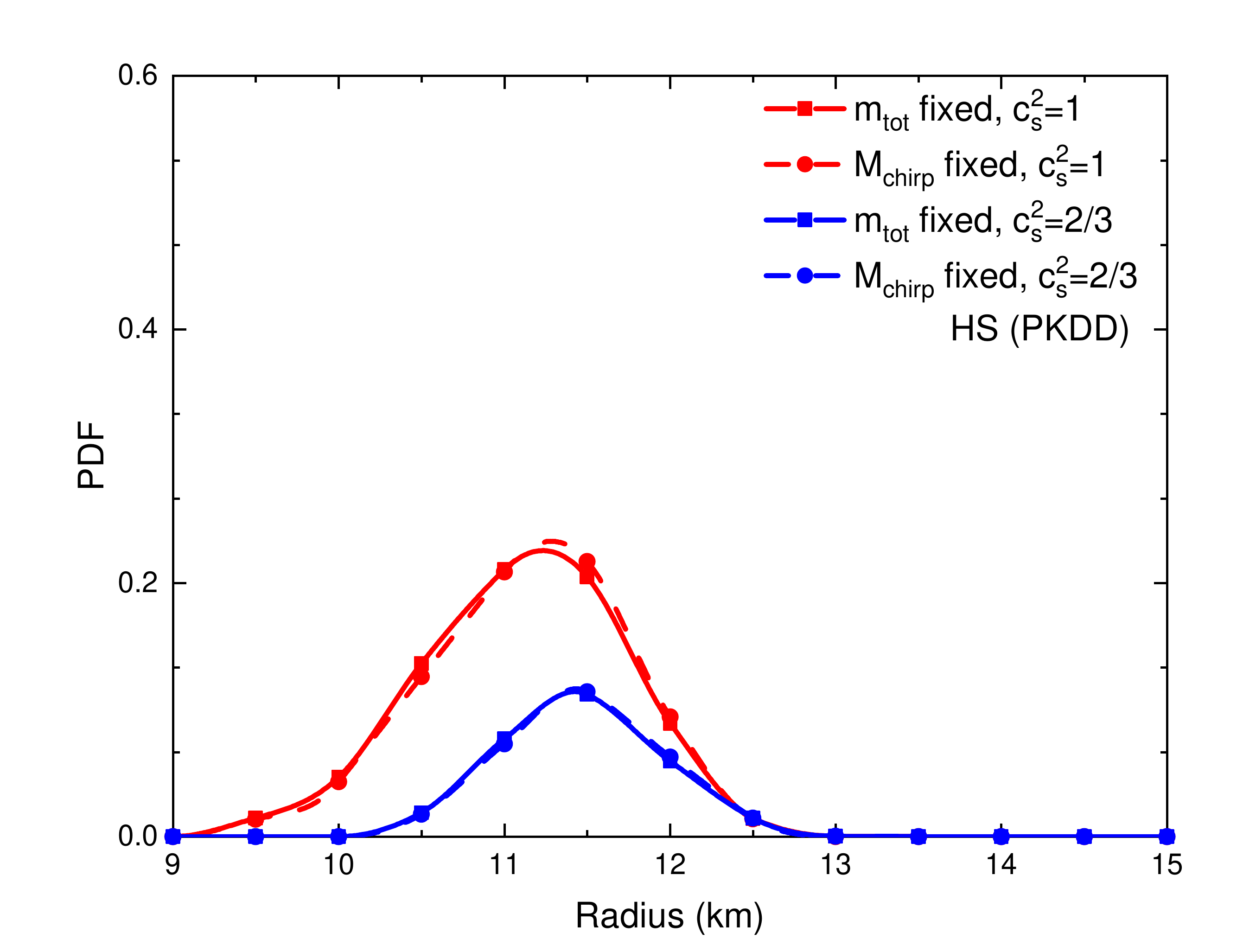}
\caption{Distribution of radius $R_{1.4}$ obtained with our analysis for the PKDD EoS and the BHS case, calculated using a constant value for $m_\mathrm{tot}$ (filled square symbol) or a constant value for $M_\mathrm{chirp}$ (filled circle symbol), where the data are extracted from Ref.~\cite{Abbott:2019}. We have considered EoS where the sound speed $c_s^2$ has been fixed to $1$ (red lines) and to $2/3$ (blue lines).
}
\label{fig:MchirpVsMtot}
\end{figure}

The fifth-order post-Newtonian correction to the wavefront phase is the tidal deformability~\cite{Flanagan:2008,Damour:2009}, which is defined as,
\begin{equation}
\tilde{\Lambda} = \frac{16}{13}\Big[\frac{(m_1+12m_2)m_1^4}{m_\mathrm{tot}^5}\Lambda_1+1\leftrightarrow2 \Big] \, .
\end{equation}

The impact of the two prescriptions for $m_1$ and $m_2$ into our analysis is shown in Fig.~\ref{fig:MchirpVsMtot}, where we present the posterior distribution for $R_{1.4}$ for the PKDD BHS case. The posteriors are calculated by employed the two prescriptions: a constant value for $m_\mathrm{tot}$ or a constant value for $M_\mathrm{chirp}$. This illustrates that the small differences in the masses $m_1$ and $m_2$ obtained using one or the other of the two prescriptions have a negligible effect for our analysis.

\begin{figure}[t]
\centering
\includegraphics[width=0.48\textwidth]{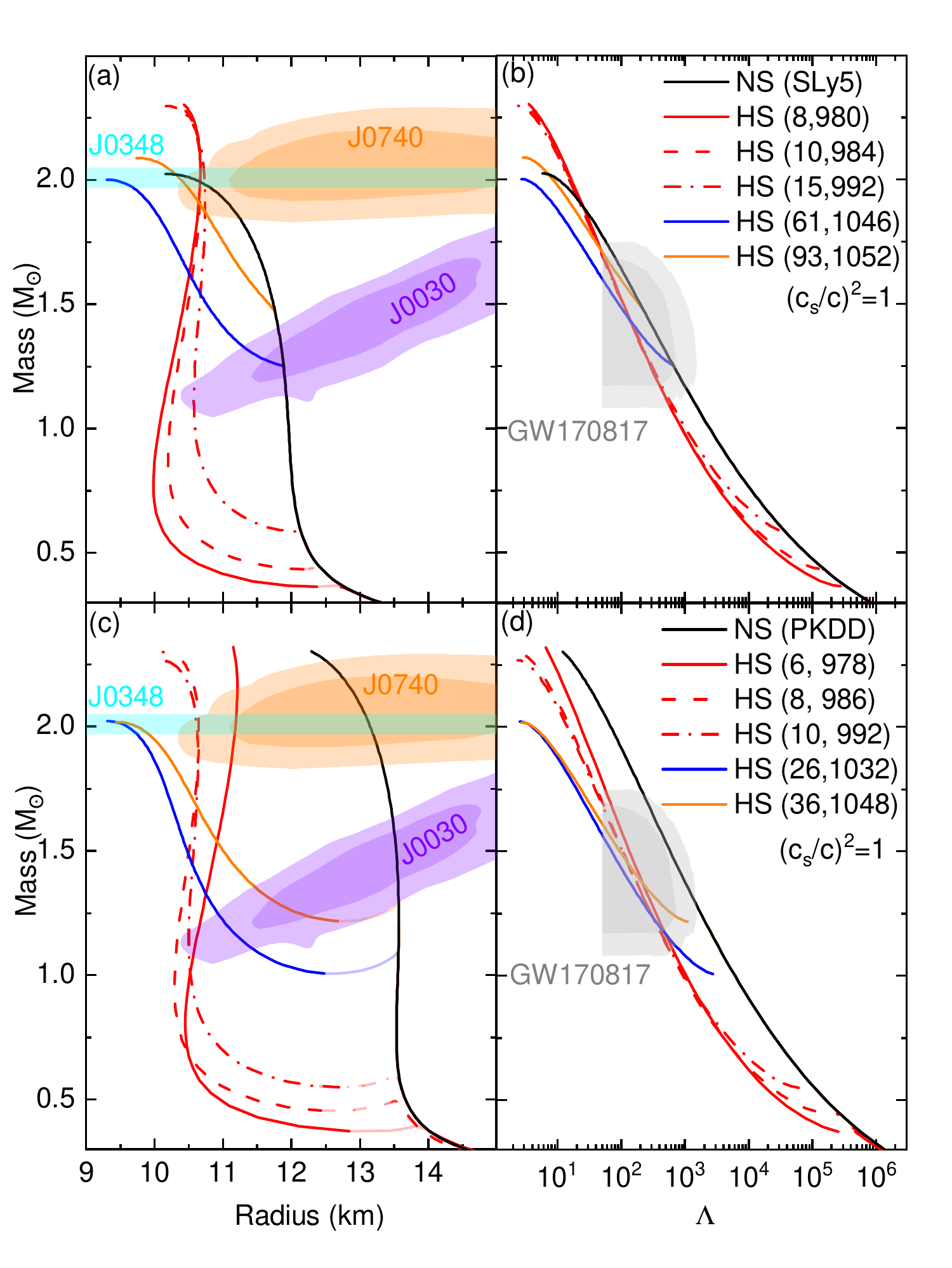}
\caption{Same as Fig.1 but the contours of GW170817 are extracted from Ref.~\cite{Abbott:2019}.}
\label{fig:MRLVC}
\end{figure}

We now analyse the impact of the extracted $\tilde{\Lambda}$-PDF in our analysis by exploring the $\tilde{\Lambda}$-PDF from another parameter estimation approach.
In Fig.~\ref{fig:MRLVC}, the contours of GW170817 is calculated from the prediction by the LVC~\cite{Abbott:2019}, instead of Ref.~\cite{De:2018}, where the mass asymmetry of the binary system is given as the 1$\sigma$ and 2$\sigma$ confidence intervals, as in Fig.~1 of the paper.
Comparing with Fig.~1 of the paper, the contour associated to GW170817 is now larger and allow a larger number of EoS.
The tension between NICER observations and GW170817 is however still visible:
the EoSs compatible with the contours from NICER observatory (PKDD for instance) have marginal overlap with the contours of GW170817.

\begin{figure}[t]
\centering
\includegraphics[width=0.48\textwidth]{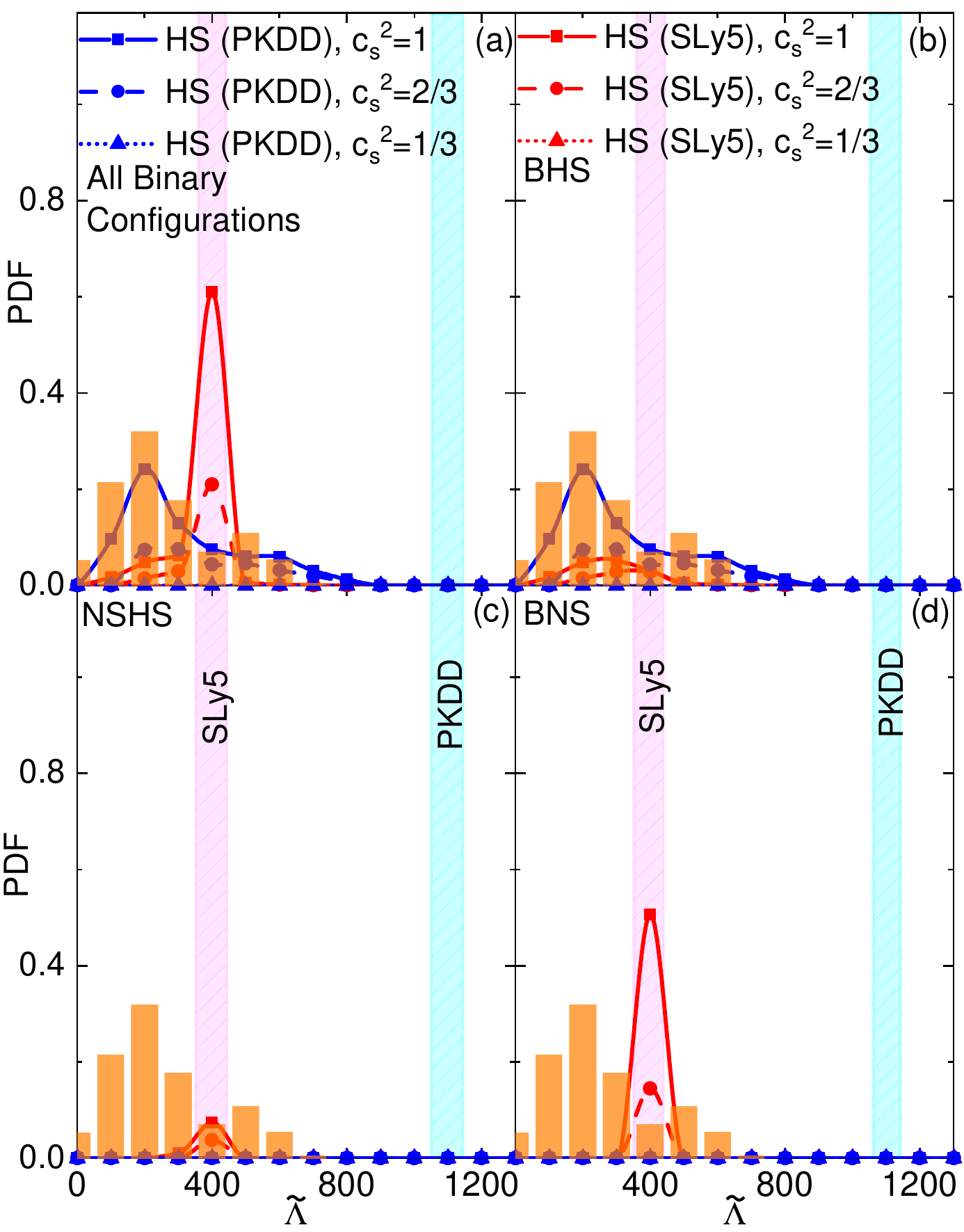}
\caption{Same as Fig.~2 but employing data from Ref.~\cite{Abbott:2019}.}
\label{fig:LambdaLVC}
\end{figure}

The posterior probability as a function of the effective tidal deformability $\tilde{\Lambda}$ from Ref.~\cite{Abbott:2019}, is shown in Fig.~\ref{fig:LambdaLVC} for BNS, BHS, NSHS configurations. 
Although, the $\tilde{\Lambda}$-PDF is qualitatively quite different between Ref.~\cite{De:2018} (see Fig.~2 in the paper) and Ref.~\cite{Abbott:2019} (with a double peaked distribution), the posteriors are almost identical between Fig.~2 of the paper and Fig.~\ref{fig:LambdaLVC}: BNS systems are preferred only with soft nuclear EoS with FOPT, NSHS systems are not favored by either soft and stiff nuclear EoSs with FOPT and finally stiff nuclear EoS with FOPT is overlapping best with data for the BHS systems as shown in Tab. I. The overlaps are given in Tab.~I in the paper, and they are very close.

\begin{figure}[t]
\centering
\includegraphics[width=0.48\textwidth]{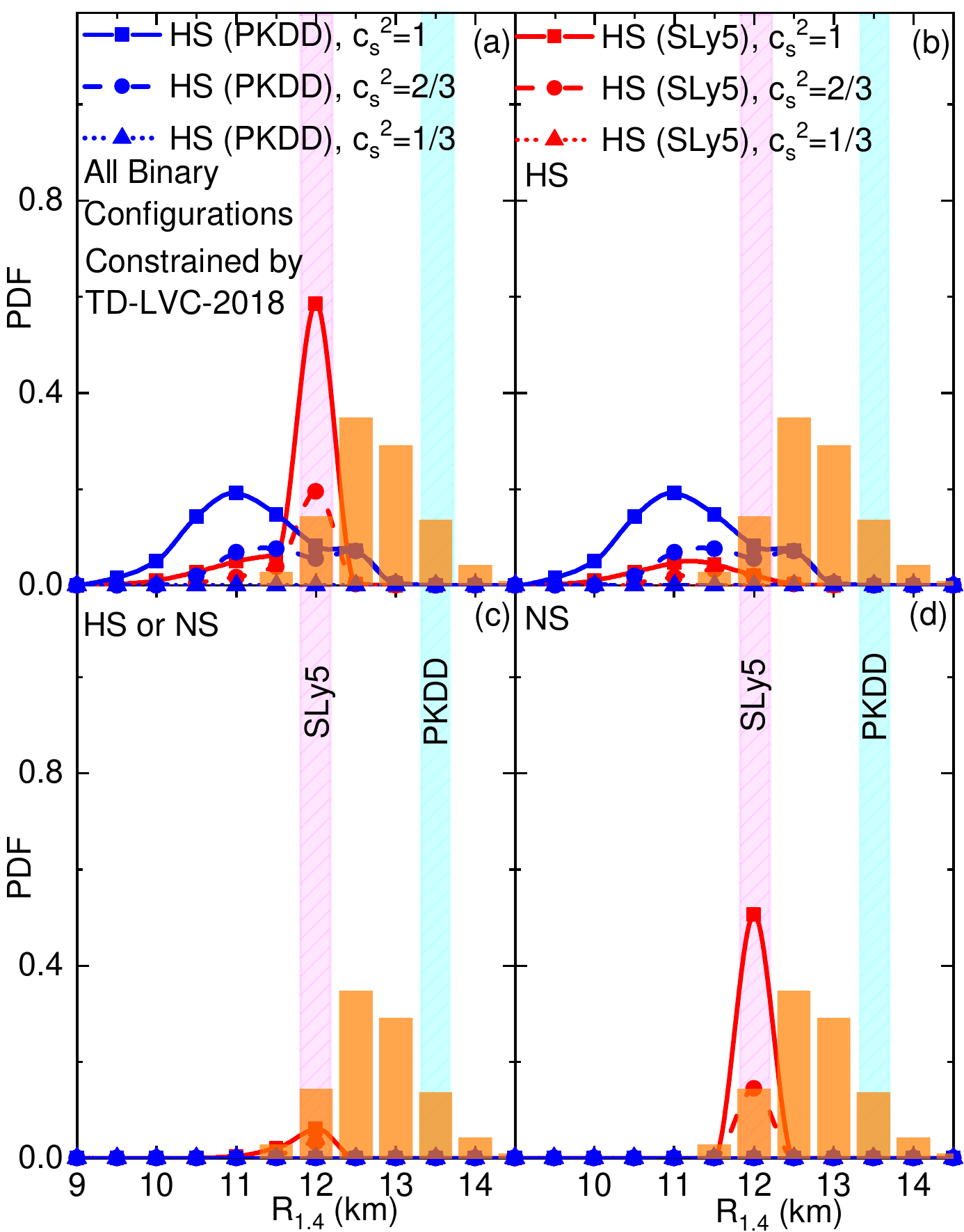}
\caption{Same as Fig.~3 but employing the $\tilde{\Lambda}$-PDF from Ref.~\cite{Abbott:2019}.}
\label{fig:radiusLVC}
\end{figure}

We now come to the predictions for the radii $R_{1.4}$, using the same approach as in the paper. Results are shown in Fig.~\ref{fig:radiusLVC} where the EoSs are now constrained by $\tilde{\Lambda}$ from Ref.~\cite{Abbott:2019}. The results are qualitatively very similar. There are however small quantitative differences which are also reported in Tab.~II of the paper. The largest difference is of the order of 300m.

\bibliography{biblio}